\documentclass[11pt]{article}
\usepackage{amssymb, amsmath}
\usepackage{color, pdfcolmk}
\usepackage{graphics}
\usepackage{graphicx}
\usepackage{amssymb}
\usepackage{epsf}
\usepackage{verbatim}

\newtheorem{theorem}{Theorem}
\newtheorem{lemma}{Lemma}

\newtheorem{definition}[lemma]{Definition}
\newtheorem{proposition}[lemma]{Proposition}
\newtheorem{corollary}[lemma]{Corollary}

\newtheorem{remark}[lemma]{Remark}

\newcommand{\eps}{\varepsilon}
\newcommand{\ie}{{i.e.}}
\newcommand{\cS}{\mathcal{S}}
\newcommand{\cL}{\mathcal{L}}

\begin{document}

\title{Extremely Chaotic Boolean Networks}

\author{Winfried Just\thanks{To whom correspondence should be addressed. E-mail: just@math.ohiou.edu}{ }\footnote{Department of Mathematics, Ohio University, Athens, OH 45701} \ \ and Germ\'an A. Enciso\footnote{Harvard Medical School, Department of Systems Biology}}

\maketitle

\begin{abstract}

It is an increasingly important problem to study conditions on the structure of a network that guarantee a given behavior for its underlying dynamical system.  In this paper we report that a Boolean network may fall within the chaotic regime, even under the simultaneous assumption of several conditions  which in randomized studies have been separately shown to correlate with ordered behavior.   These properties include using at most two inputs for every variable, using biased and
canalyzing regulatory functions, and restricting the number of negative feedback loops.

We also prove for $n$-dimensional Boolean networks that if in addition the number of outputs for each variable is bounded and there exist periodic orbits of length $c^n$ for $c$ sufficiently close to 2, any network with these properties must have a large proportion of variables that simply copy previous values of other variables.  Such systems share a structural similarity to a relatively small Turing machine acting on one or several tapes.

\end{abstract}


The concept of a \emph{Boolean network} was originally proposed in the late 1960's by Stuart Kauffman to model gene regulatory behavior at the cell level \cite{Kauffman:N1969,Kauffman:JTB1969}.   This type of modeling can sometimes capture the general dynamics of continuous systems in a simplified framework without the choice of specific nonlinearities or parameter values; see for instance~\cite{Albert:JTB2003}.   Boolean networks are used in several other disciplines such as electrical engineering, computer science, and control theory,
and analogous definitions are known under various names such as sequential dynamical systems
\cite{Laubenbacher:A2001} or Boolean difference equations \cite{GhilI}.

In studying the dynamics of Boolean networks, Kauffman distinguishes an \emph{ordered regime} and a \emph{chaotic regime} and argues that the dynamics of actual gene regulatory networks should be close to the boundary of these two regimes (see \cite{origins}
 for a review).
  Each regime is characterized by several hallmark properties that usually, but not always, are present simultaneously. In this paper we focus on one such property, the existence of exponentially long orbits, although two additional such properties will be briefly considered as well.

 Since the state space of an $n$-dimensional Boolean system is finite, each trajectory must eventually reach a periodic orbit of length $\leq 2^n$ or a fixed point.  The ordered regime is characterized by relatively short orbits whose length scales like a low-degree polynomial in~$n$.  In contrast, orbits whose length scales exponentially in~$n$ are a hallmark of the chaotic regime.  We will call an $n$-dimensional Boolean network
 \emph{$c$-chaotic} if it has an orbit of length $\geq c^n$, for $1 < c < 2$.

An $n$-dimensional \emph{Boolean dynamical system} or \emph{Boolean network}
consists of $n$ variables $s_1,\ldots, s_n$, each of which can have value 0 or 1 at any given time step $t$.  The updates for each variable are calculated by $s_i(t+1) = g_i(s_1(t),\ldots,s_n(t))$, where $g_i$ is called the $i$-th  \emph{regulatory function} of the system (taking our motivation from Boolean models of gene regulatory networks).   In the study of \emph{random Boolean networks (RBNs),} these regulatory functions are randomly and independently drawn from a specified distribution, and the dynamics of the resulting network is simulated for a sample of initial states.

Much attention in empirical studies has focused on studying which properties of the regulatory functions correlate with dynamics within the ordered regime.   Already in his 1969 papers \cite{Kauffman:N1969,Kauffman:JTB1969}, Kauffman focused  his attention on Boolean networks where each $g_i$
 depends only on a bounded number of inputs, regardless of the dimension
  of the network.
This corresponds to findings about actual gene regulatory networks which show that most genes are directly regulated by a small number of proteins in a scale-free manner \cite{ArnoneDavidson,Tong:Science2004}. We also note that the scale-free distribution implies that
in some large subnetworks each variable has a bounded number of \emph{outputs,} \ie, acts as input only for a bounded number of variables.

A so-called \emph{NK-network} is an RBN of dimension $N$ obtained by randomly choosing a set of $K$ inputs for each regulatory function, and then choosing $g_i$ randomly from the uniform distribution of all Boolean functions on this set of inputs.  The choices for different $i$ are independent.
Since not all Boolean functions on $K$ variables depend on all inputs, one can consider an NK-network as a random network where each regulatory function takes at most~$K$ inputs.

For $K = 2$ the dynamics of $NK$-networks tends to be in the ordered regime;
in particular, the median length of orbits is on the order of $\sqrt{N}$.  In contrast, when
 $K > 2$, the dynamics  tends to be chaotic \cite{origins}.

However, several additional restrictions on the $g_i$'s still tend to result in RBNs with ordered dynamics, even for large average number of inputs.

The \emph{bias}  $\Lambda$  of a
Boolean function is the fraction of input vectors for which the function outputs 1.  Studies of RBNs in which each $g_i$ has bias $\Lambda$ close to~0 or~1 show that the dynamics tends to be in the ordered regime even if the $g_i$'s have a relatively large numbers of inputs \cite{DerridaStauffer, WeisbuchStauffer}.

A Boolean function $g_i$ that depends on variables $x_1, \ldots , x_\ell$ is
\emph{canalyzing}
if there exist one input variable $x_c$ and Boolean values $u, v$  such that
$g_i(x_1, \ldots , x_\ell) = v$ whenever $x_c = u$.  A stronger property is the notion of a \emph{nested canalyzing function.} Empirically characterized Boolean regulatory functions tend to be nested canalyzing \cite{Harris}. Since for Boolean functions with at most two inputs the two notions coincide, we will not define this stronger property here.  RBNs in which all regulatory functions are nested canalyzing functions were found to have dynamics in the ordered regime, even though individual $g_i$'s may have numerous inputs \cite{nestcan}.

Finally, RBNs with no or only few negative feedback loops  tend not to reach long orbits, even when the $g_i$'s are not restricted to those with the properties listed above \cite{Sontag:Laubenbacher}.  This behavior can be compared to that of \emph{continuous} systems with no negative feedback, which are well known to converge generically towards an equilibrium (\cite{Smith:monotone} and see the next section).

Simulation studies of RBNs can only demonstrate that exponentially long orbits are not reached from the initial conditions that are sampled.
Our research was guided by the following question: under what conditions for the network can the absence of $c$-chaos for $c$ sufficiently close to 2 be rigorously proved?  In particular, we were interested in whether a combination of the conditions that were known empirically to generate RBNs with ordered dynamics would preclude the existence of very long orbits.
In this paper we report that even when all these assumptions are made simultaneously, for every positive $c < 2$ one can construct examples of Boolean networks whose dynamics exhibits $c$-chaos. This is true even when the number of outputs per variable is limited to 2.  However, the situation changes somewhat if we assume in addition to the latter that all, or a specified proportion of the regulatory functions take \emph{exactly} two inputs:  for such systems it is possible to prove the absence of $c$-chaos for some $c < 2$.

Boolean systems in which most regulatory functions take only one input share a structural stability with a small Turing machine that acts on one or several tapes.  We conclude that this Turing-like structure is the only possibility for building some types of extremely chaotic dynamics into  Boolean systems from a certain class.

\section{Major Results}

We define a $(b,r)$-Boolean system as a system in which each regulatory function $g_k$ has at most $r$ inputs, and each variable has at most $b$ outputs.  If $r = 2$, we call the system \emph{quadratic;} a $(2,2)$-system is called \emph{bi-quadratic.} A regulatory function that depends on only one variable is called \emph{monic;} a non-monic quadratic regulatory function is called \emph{strictly quadratic.}
A Boolean network with only quadratic regulatory function will be called a \emph{strictly quadratic network;} a strictly quadratic bi-quadratic network will be called \emph{strictly bi-quadratic,} even if some variables have fewer than two outputs.

In the context of continuous dynamical systems, a system without negative feedback loops is called \emph{monotone} \cite{Sontag:mono, Smith:monotone}.  Special cases of monotone systems are \emph{cooperative systems} in which there are no direct inhibitory interactions
 between any two variables. Monotone and cooperative systems have been used as a modeling tool for gene regulatory systems, e.g. in \cite{Pommerening:2003, Angeli:PNAS2004, Leenheer:2006}.
While negative feedback tends to generate oscillatory dynamics, the assumption of monotonicity in continuous systems ensures, under mild additional assumptions, that a generic solution of a monotone dynamical system must converge towards an equilibrium.  In contrast, cooperative Boolean systems can still have exponentially long orbits (see e.g. \cite{Sontag:mono:Arxiv2007}, \cite{Just:Enciso:embedding}).

Cooperative Boolean  systems have regulatory functions that can be expressed using only AND and OR gates, \ie, with no use of negations.  This can be seen by considering the disjunctive normal form of the Boolean maps.
In particular, the only non-constant regulatory functions allowed in quadratic cooperative Boolean systems must have the form
$g_k = s_{i_k} \wedge s_{j_k}$, $g_k = s_{i_k} \vee s_{j_k}$, or $g_k = s_{i_k}$.
Note that if $g_k$ is constant then we get identical dynamics along attractors if we replace it with the monic function $g_k = s_k$.
Since transient states are irrelevant for our results, we will without loss of generality assume that all regulatory functions are non-constant.

All regulatory functions that are allowed in quadratic cooperative Boolean systems are canalyzing, and the two
permissible strictly quadratic regulatory functions have bias $\Lambda = 0.25$ and $\Lambda = 0.75$ respectively.
Our first theorem shows that even if all regulatory functions have at most two inputs, are strongly biased and canalyzing, and negative feedback is totally absent, the system may still have exponentially long periodic orbits.

\begin{theorem} \label{theorem counterexample}  Let $c, c_1$ be constants with $1 < c < 2$ and $1 < c_1 < 10^{1/4}$.  Then for all sufficiently large $n$ there exist
$n$-dimensional cooperative Boolean networks that are, respectively:

\noindent
(i)  bi-quadratic and $c$-chaotic,

\noindent
(ii) strictly quadratic and
$c$-chaotic,

\noindent
(iii) strictly bi-quadratic and
$c_1$-chaotic.
\end{theorem}

Moreover, as we we will show in Theorem~\ref{thm+} of Appendix~A, the construction in point~(i) can be done in such a way that
\begin{enumerate}
 \item the values of most of the variables of the system continue to alternate between 0 and 1 over time, for most initial conditions, and
 \item with probability arbitrarily close to one, changing the value of a randomly chosen variable in almost every initial condition sends the trajectory of the system into a different basin of attraction.
\end{enumerate}
We will consider these properties in more detail in the Discussion Section below.   This confirms that the system displays several hallmarks of extremely chaotic behavior as generally defined in the literature.

\bigskip

Our second major result shows that
it is not possible, for $c$ less than but arbitrarily close to $2$, to construct $n$-dimensional bi-quadratic cooperative Boolean networks with $c$-chaotic dynamics in such a way that all or even a given proportion of regulatory functions are strictly quadratic.
 This result has an interesting interpretation from the point of view of theoretical computer science.
Consider a sequence of variables $k_1, \ldots , k_m$ such that $g_{k_{i+1}} = s_{k_i}$ for all $i \in \{1, \ldots , m-1\}$.  The dynamics of the system on these
variables is analogous to that of a memory tape of a Turing machine that advances by one position at each time step.  A new value may be written to position $k_1$ at each time step, and this value may be read $\ell$ time steps later by some regulatory function off position $k_{\ell + 1}$.  If $k_m = k_1$, the tape is
`read-only,' and a constant regulatory function can be considered a special case of a `read-only' tape of length~one.
A tape could split into two or more branches,
but the values on these branches would eventually be only copies of each other.
Thus any cooperative system that contains monic regulatory functions can be conceptualized as a Turing machine whose internal states correspond to
all non-monic variables and that acts
on one or more tapes, possibly branching or of varying lengths.  Let us call an $n$-dimensional Boolean system  an \emph{$(M, n)$-Turing system} if at least $n - M$ of the regulatory functions are monic. While every $n$-dimensional Boolean system is an $(n,n)$-Turing system in the sense of the above definition,  if $M < n$, the roles of the `machine' and the `tapes' can be neatly separated.
If also monic regulatory functions
$g_k = \neg s_{i_k}$ may occur in the system, then the connection with the Turing machine metaphor becomes more tenuous, but for convenience we will still use this terminology even if the system is not assumed cooperative.

The idea of the proof of Theorem~\ref{theorem counterexample}(i) sketched below is based on the metaphor of a Turing machine.  The systems constructed in this proof are $(M(n),n)$-Turing systems such that $\lim_{n \rightarrow \infty} \frac{|M(n)|}{n} = 0$.  While the metaphor of a Turing machine acting on one or several tapes
readily comes to mind as a mechanism for constructing counterexamples,
it is far from obvious whether totally different systems with analogous properties might exist.  But Theorem~\ref{GenTthm}  below implies that for $c$ sufficiently close to $2$ our construction is in some sense the only possibility to build $c$-chaos into certain systems.

We say that a Boolean system $(\Pi, g)$ is \emph{$\eps$-biased} if every non-monic regulatory function has bias $\Lambda$ with
  $|\Lambda - 0.5| \geq \eps$.  In particular, quadratic Boolean systems are $0.25$-biased.
  Cooperative Boolean systems with regulatory functions that can take three or more inputs need not be $\eps$-biased for any
  $\eps > 0$. For example, $\Lambda = 0.5$ for the Boolean function with three input variables that takes the value $1$ iff the majority of input variables are equal to~$1$.

\begin{theorem}\label{GenTthm}
Let $\varepsilon, \alpha > 0$ and let $b, r$ be positive integers. Then there exists a positive constant $c(\eps, \alpha, b, r) < 2$ such that for every $c > c(\eps, \alpha, b, r)$ and sufficiently large $n$,
 every $c$-chaotic, $n$-dimensional $\eps$-biased $(b,r)$-Boolean system is an  $(\alpha n, n)$-Turing system.
\end{theorem}

A canalyzing Boolean function has bias $\Lambda = 0.5$ iff it is monic.  Since there are only finitely many
Boolean functions on any fixed number of inputs, the conclusion of Theorem~\ref{GenTthm} will hold in particular for all $(b,r)$-Boolean systems in which all regulatory functions are canalyzing.

Theorems~\ref{theorem counterexample} and~\ref{GenTthm} combined show that while the assumptions of  canalyzing regulatory functions and absence of negative feedback do not impose a nontrivial bound on the lengths of orbits in $(b,r)$-Boolean systems, the assumption that all or sufficiently many regulatory functions be sufficiently biased does impose such bounds.

\section{Sketches of the Proofs}

Here we sketch the proofs of our main theorems. Detailed proofs were first reported in~\cite{PartI} and~\cite{PartII}; improved versions of these proofs are given in Appendices~A and~B.

\subsection{Proof of Theorem~\ref{theorem counterexample}}

The idea of Theorem~\ref{theorem counterexample}(i) is based on thinking of variables
$s_{1}, \ldots , s_N$
as the internal states of a
Turing machine that writes successive binary codes of integers $0, \ldots , 2^{N} - 1$ to the variables that are organized in a long circular tape.  A straightforward implementation would have $s_i(t):=s_{i+1}(t-1)$ for $i=1,\ldots, N-1$ and an internal variable
$\mbox{\it mode}$
 such that if $mode=\mbox{\it{rotate}}$, then
$s_N(t):=s_1(t-1)$, and if $\mbox{\it mode}=\mbox{\it{switch}}$, then $s_N(t):=\neg s_1(t-1)$.

For instance, if $N=7$ and  $(s_N,\ldots,s_1)=(0\ 1\ 0\ 0\ 1\ 1\ 1)$ at $t=0$, then letting $\mbox{\it mode}=\mbox{\it{rotate}}$ for three time steps and $mode=\mbox{\it{switch}}$ for another four, one reaches the new state $(s_N,\ldots,s_1)=(0\ 1\ 0\ 1\ 0\ 0\ 0)$ at $t=N$, which corresponds to adding one to the binary code for $t=0$.
After every $N$ time steps one necessarily reaches the binary code of the successor (modulo $2^N$) of the previously coded integer, which guarantees for the overall system an orbit of length at least $2^N$.

The major problem with this implementation is that it involves negation and thus is non-cooperative.
Our construction overcomes this obstacle in the following way.
Instead of the binary digits $s_i$, let each of the variables $S_1,\ldots,S_K$ be a binary sequence of length $L$, and let $N=KL$.  Importantly, the values of $S_i$ are not arbitrary but chosen from the image of an injective function $\Gamma:\{0, \ldots , 2^\ell - 1\} \to 2^{\{1,\ldots, L\}}$, i.e.  they are thought of as coding integers from 0 to $2^\ell - 1$.
Additionally, the values of  $\Gamma(x)$ are required to have exactly $L/2$ nonzero entries.
 Such a function $\Gamma$ exists for suitable choices of $\ell$ and $L$.
 Again, let $S_i(t):=S_{i+1}(t-1)$ for $i=1,\ldots, K-1$.  Similarly, if  $\mbox{\it mode}=\mbox{\it{rotate}}$, then
$S_K(t):=S_1(t-1)$.  If $\mbox{\it mode}=\mbox{\it{switch}}$, then $S_K(t)$ is the code for the integer $\Gamma^{-1}(S_1) + 1$, where addition is defined modulo $2^\ell$.   The variable $\mbox{\it mode}$ will be coded by a binary string of length~$2$, with $(0,1)$ standing for $\mbox{\it rotate}$ and $(1,0)$ standing for $\mbox{\it switch}$.

\begin{figure}[ht]
\centerline{\includegraphics[width=3.3in]{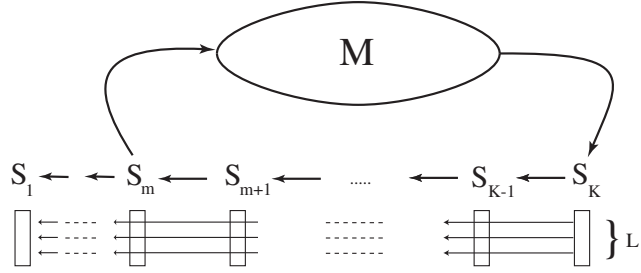} }
\caption{Architecture of the Boolean network constructed in this proof. }
\label{figure Fig1}
\end{figure}

Now consider the Boolean vector function that takes as inputs the variables $S_1(t-1)$ and $\mbox{\it mode}(t-1)$ and outputs
$S_K(t)$ and $\mbox{\it mode}(t)$ --- this function constitutes the computing core of the system, and one can refer to it as the `Turing machine' $M$ within the Boolean network.   The coding function $\Gamma$ is defined in such a way that the core function in $M$ can be implemented without the use of negation.
Allowing for some delay in the output, such a machine in turn can be coded by one that uses only binary AND- and OR- gates.  The resulting delay poses another technical problem since $\mbox{\it mode}(t)$ depends on  $\mbox{\it mode}(t-1)$;
Lemma~\ref{lemma switch} of Appendix~A gives our solution of this problem.
At every time $t$ we use up one of the $K$ states $S_1,\ldots,S_K$ to mark the beginning of the encoded information. Thus we obtain a bi-quadratic cooperative Boolean system with orbits of length at least
$2^{\ell(K-1)} = 2^{\ell(N - L)/L}$, and the number of internal variables of the Turing machine $M$
 is a number $T(L)$ that depends only on $L$.
Figure~\ref{figure Fig1} illustrates our construction.  The `Turing machine' $M$ corresponds to the union of the subnetworks $B$ and $D$ of Figure~\ref{figure cooperative} of Appendix~A.

For a given $1 < c < 2$
 we can find positive integers $\ell < L$ such that
$\binom{L}{L/2} > 2^\ell > c^L$ and
 $2^{\ell(N - L)/L} > c^N$.  The total number of variables in the system is given by $n = N + T(L)$,
and it follows that for sufficiently large $n$, we can choose $\ell, L, N$ so that the system we constructed will contain an orbit of length $> c^n$, as stated in Theorem~\ref{theorem counterexample}(i).

For the proof of part~(ii) let $(\Sigma, f)$ be a quadratic cooperative Boolean system  of dimension $n-2$ that contains an  orbit of length $c^n$.   Let $\Pi = \{0, 1\}^{n}$, let $g_k = f_k$ whenever $k < n-1$ and $f_k$ is strictly quadratic, let
$g_k = s_{i_k} \wedge s_n$ whenever $k < n$ and $f_k = s_{i_k}$, and let $g_{n-1} = g_n = s_{n-1} \wedge s_n$.  Then $(\Pi, g)$ is cooperative, quadratic, and has only strictly quadratic regulatory functions.
For a state $s \in \Sigma$ define a state  $s^* \in \Pi$ by
$s^* = (s_1, \ldots, s_{n-2}, 1, 1)$.  Then the orbit of $s^*$ in $(\Pi, g)$ has the same length as the orbit of
$s$ in $(\Sigma, f)$.

The proof of part~(iii) is given in Appendix~A.

\subsection{Proof of Theorem~\ref{GenTthm}}

The proof is based on the observation that very large subsets of the state space of Boolean systems must be \emph{balanced} in the following sense. Let $\cS = \{s^\ell: \, \ell \in \cL\}$
be a subset of the state space $\Pi = \{0,1\}^{n}$.
Consider an  $r$-element subset $I = \{i_1, \ldots , i_r\}$ of $\{1, \ldots , n\}$ with
$i_1 < \dots < i_r$, and let $\sigma : \{1, \ldots , r\} \rightarrow \{0,1\}$.
Define a ratio
$\xi_{I}^{\sigma}(\cS)$ by

$$\xi^{\sigma}_{I} (\cS) = \frac{|\{\ell \in \cL: \, \forall u \in \{1, \ldots ,r\} \ s_{i_u}^\ell = \sigma(u)\}|}{|\cL|}.$$

If $\cS$ is randomly chosen, then $\xi^{\sigma}_I(\cS)$ will be close to $2^{-r}$.  Intuitively, a balanced set $\cS$ is one in which
only few of the ratios $\xi_I^{\sigma}$ differ substantially from $2^{-|I|}$.
More precisely, if $\beta, \gamma > 0$, then we will say that $\cS$ is \emph{$\beta$-$\gamma$-$r$-balanced} if for every family $P$ of pairwise disjoint
subsets $I$ of $\{1, \ldots ,n\}$ with $|\bigcup P| \geq \beta n$ and $1 \leq |I| \leq r$ for each $I \in P$ there exists $I \in P$ such that
$|\xi^\sigma_I(\cS) - 2^{-|I|}| < \gamma$ for all relevant $\sigma$.

In the first part of the proof of Theorem~\ref{GenTthm} we show that
for any given
 positive integer $r$, and $\beta, \gamma > 0$ with $\gamma < 2^{-r}$ there exists a constant $c < 2$ such that
 for sufficiently large~$n$, every subset $\cS$ of $\{0,1\}^{n}$ of size $\geq c^n$ is  $\beta$-$\gamma$-$r$-balanced.
The proof of this fact uses the probabilistic method.  We fix $c < 2$ and derive an upper bound for the probability that a randomly chosen subset of size $\geq c^n$ of $\{0,1\}^{n}$ is \emph{not} $\beta$-$\gamma$-$r$-balanced.  For $c$ sufficiently close to $2$, this probability will be less than $\binom{2^n}{c^n}^{-1}$.  But if there exists \emph{any} unbalanced subset of $\{0,1\}^{n}$ of this size,
then it would be picked with probability at least $\binom{2^n}{c^n}^{-1}$, which leads to a contradiction.

Now consider an $\eps$-biased, $n$-dimensional $(b,r)$-Boolean system with a $\beta$-$\gamma$-$r$-balanced orbit $\cS$.  For suitable choices of $\beta, \gamma$, there will be a subset $S^*$ of $S$ with
$|\cS^*|/|\cS| \approx 1$ and a set $J \subseteq \{1, \ldots ,n\}$ of size $|J| < (r+1)\beta n$ so that $\xi_I^\sigma(\cS^*) = 2^{-|I|}$ for all relevant $\sigma$
whenever $I$ is the set of inputs of a variable $k \notin J$ and $I$ is disjoint from $J$. Let $k$ be a variable outside of $J$ whose regulatory function is biased; wlog assume $\Lambda \geq 0.5 + \eps$.  If the set of inputs $I$ of $g_k$ is disjoint from $J$, then
$$\xi_{k}^1(\cS) \geq \frac{|\{\ell: s^\ell \in \cS^* \ \& \ s^\ell_k = 1\}|}{|\cS|} \geq \frac{(0.5 + \eps)|\cS^*|}{|\cS|},$$
and by the choice of $J$ we get a contradiction with the assumption that $S$ was  $\beta$-$\gamma$-$r$-balanced and thus in particular
$\beta$-$\gamma$-$1$-balanced.

\section{Summary and Discussion}\label{section discussion}

Exponentially long orbits of Boolean systems are a hallmark of the chaotic regime.
In empirical studies of RBNs very long orbits tend not to be reached
when the number of inputs for each regulatory function is bounded by 2
\cite{origins}, when the regulatory functions are strongly biased \cite{DerridaStauffer, WeisbuchStauffer}, when all regulatory functions are nested canalyzing functions \cite{nestcan}, or when there are few negative feedback loops \cite{Sontag:Laubenbacher}.

Theorem~\ref{theorem counterexample}(ii) shows that even the
conjunction of these four conditions is not sufficient to prove a nontrivial upper bound on the lengths of possible orbits.  If in addition an upper bound on the number of outputs per variable is assumed, then nontrivial upper bounds can be derived (Theorem~\ref{GenTthm}).  Such bounds can be derived assuming only restrictions on the number of inputs and outputs per variable and that a given fraction of regulatory functions is sufficiently biased.  Even with these additional assumptions, exponentially long orbits may exist (Theorem~\ref{theorem counterexample}(iii)). Without assumptions on the fraction of sufficiently biased regulatory functions, no nontrivial upper bound on the lengths of orbits can be derived for bi-quadratic cooperative Boolean systems (Theorem~\ref{theorem counterexample}(i)).  For $c$ sufficiently close to~$2$, all $n$-dimensional bi-quadratic Boolean systems with orbits of length $c^n$ must be structurally similar to a Turing machine acting on one or several tapes (Theorem~\ref{GenTthm}).

Theorem~\ref{GenTthm} has yet another alternative interpretation.
Variables with monic regulatory functions  just record the values of other variables (in some cases their negations, if cooperativity is not assumed) at a certain time in the past.
 Thus if we allow time delays in the definitions of regulatory functions, we can remove all but the
first variable on each `tape' and define a \emph{Boolean delay system} on the remaining variables that will have equivalent dynamics, in particular, that will have  orbits of the same length as the original system.  Gene regulation always involves a delay between gene transcription and the time when the translated
gene product becomes available as a regulator, such as a transcription factor. Boolean delay systems with internal variables that record the state of other variables were proposed as models of gene regulatory networks in the framework of `kinetic logic'  by R.~Thomas \cite{ThomasI, ThomasII}.
Internal variables are not needed in the framework of continuous-time Boolean delay systems as
studied in \cite{GhilI, GhilII}.
See \cite{GhilIII} for a comprehensive survey and additional references.  Our $(M,n)$-Turing systems can be conceptualized in this framework  as continuous-time Boolean delay systems with $M + m$ Boolean variables and rational delays, where $m$ is the number of read-only tapes (see Appendix~F).
Thus Theorem~\ref{GenTthm} implies that for any given
 for $\alpha, \eps > 0$ and positive integers $b,r$ there exists a positive constant  $c < 2$ such that for sufficiently large $n$, every $n$-dimensional $\eps$-biased $(b,r)$-Boolean system with an orbit of length at least $c^n$ is equivalent to a Boolean delay system with rational delays and at most $\alpha n$ Boolean variables.

 For given $\eps, \alpha, b, r$ let $c(\eps, \alpha, b, r)$ denote the largest constant $c$ for which the conclusion of Theorem~\ref{GenTthm} holds.
For example, if a bi-quadratic cooperative system of sufficiently large dimension $n$ has an orbit of length $> c(0.25, 0.1, 2, 2)^n$, then
at least $90\%$ of all regulatory functions must be monic; if such a system has an orbit of length $> c(0.25, 1, 2, 2)^n$, then
at least some of the regulatory functions must be monic. Our proof of Theorem~\ref{GenTthm} gives upper bounds for $c(\eps, \alpha, b, r)$.  Numerical explorations show that for small $b,r$ and relevant~$\eps$ the dependence on $\alpha$ of this upper bound is almost perfectly linear (see Figure~\ref{figure cees} of Appendix~C). In particular, for the case of bi-quadratic cooperative systems, when $\eps = 0.25$ and $b = r = 2$, we get the following linear approximation  of the
upper bound: $c(0.25, \alpha, 2, 2) \leq   2 - \, 0.0041\alpha$.  For $\alpha$ sufficiently close to 1, we were able to improve this upper bound to $c(0.25, \alpha, 2, 2) \leq 10^{(2-\alpha)/4}$ (Corollary~\ref{newupper} of Appendix~C).  On the other hand, our proof of
Theorem~\ref{theorem counterexample}(iii) gives the lower bound $10^{\alpha/4}2^{1-\alpha} \leq c(0.25, \alpha, 2, 2)$ (Proposition~\ref{1(iii)prop} of Appendix~A).  For $\alpha = 1$ the upper and lower bounds coincide, and we conclude that $c(0.25, 1, 2, 2) =10^{1/4} \approx 1.7783$ (Theorem~\ref{c22lem} of Appendix~C).  Unfortunately, the proof of the latter two results does not easily generalize to cases when $b \neq 2$ or $r \neq 2$.
It will be an interesting direction for future research to find improved estimates of $c(\eps, \alpha, b, r)$.

Let us conclude with a brief discussion of two other hallmarks of the chaotic regime. In a typical network with ordered dynamics, along the attractors reached from most initial states, a large proportion of the variables will never change their values; such variables are usually called \emph{frozen} \cite{origins}.  Let us consider a corresponding hallmark for highly chaotic systems and call a Boolean network \emph{$p$-fluid} if for a randomly chosen initial state with probability at least $p$ the network will reach an attractor along which a proportion of at most $1-p$ of the variables are frozen.

In the ordered regime, most single-bit flips in most initial conditions will leave the trajectory in the same basin of attraction.
 This property is called \emph{high homeostatic stability} in \cite{origins}.  In contrast, chaotic systems are characterized by low homeostatic
 stability.  Let us call a Boolean system \emph{$p$-unstable} if a random bit flip in a randomly chosen initial state with probability at least $p$ moves the trajectory into the basin of attraction of a different attractor.

For any given positive probability $p< 1$ and sufficiently large $n$, one can  construct systems as in  Theorem~\ref{theorem counterexample}(i) that are $p$-fluid and  $p$-unstable (Theorem~\ref{thm+} Appendix~D). Thus, the systems as in part~(i) of Theorem~\ref{theorem counterexample} can in a sense be maximally chaotic according to all three criteria considered here.

  It is also quite easy to construct strictly bi-quadratic, cooperative $1$-unstable Boolean networks of dimension $2n$ for any $n$ (see Proposition~\ref{instex} of Appendix~E).
However, we were able to prove that sufficiently high-dimensional cooperative, strictly quadratic Boolean systems cannot, for example, be simultaneously $0.9$-unstable and $1.85$-chaotic (see Theorem~\ref{pcthm}  of Appendix~E for a more general result).
This is yet another indication that extreme chaos is possible only in Turing systems.
 The result also shows that different hallmarks of the chaotic regime show quite different sensitivity to the conditions on the network architecture that were considered in this paper.

\section*{Acknowledgments}
We thank Eduardo Sontag for bringing this research topic to our attention, Xiaoping A. Shen for valuable comments, and Andrew Oster for help with illustrations.
This material is based upon work supported by the National Science Foundation
under Agreement No. 0112050 and by The Ohio State University.

\section{Appendix A: Proof of Theorem~\ref{theorem counterexample}}

A proof of  Theorem~\ref{theorem counterexample} was reported in~\cite{PartI}.  Here we include a somewhat improved version of this proof.

We associate a directed graph $D$ with vertex set $[n] :=\{1, \ldots , n\}$ with an $n$-dimensional Boolean system
$(\Pi, g)$ as follows.  A pair $<i, j>$ is in the arc set of $D$
 iff there exist states $s, r \in \Pi$ such that $s_i < r_i$ and $s_k = r_k$ for all $k \neq i$ with the property that
 $(g(s_i))_j < (g(r_i))_j$. Note that the system is bi-quadratic if both the indegree and the outdegree of all vertices
 in $D$ is at most 2.

 We will construct the systems in the proof of part~(i) of Theorem~\ref{theorem counterexample} in such a way that the associated digraph $D$ is strongly connected. This is of interest in connection with the results in \cite{Just:Enciso:embedding}.  There,
we define a local version $D_s$ of $D$ for every state $s$ as follows: A pair $<i, j>$ is in the arc set of $D_s$
 iff there exist a state $r \in \Pi$ such that either $s_i < r_i$ while $s_k = r_k$ for all $k \neq i$, and we have
 $(g(s_i))_j < (g(r_i))_j$; or $r_i < s_i$ while $s_k = r_k$ for all $k \neq i$, and we have
 $(g(r_i))_j < (g(s_i))_j$.  It is shown that if $X$ is an orbit of an $n$-dimensional cooperative Boolean system
 such that $D_s$ is strongly connected \emph{for every} $s \in X$, then $|X| \leq n$ (Theorem~25 of \cite{Just:Enciso:embedding}).
 The construction presented here shows that the analogous global property of the digraph $D$ does not impose any nontrivial bounds on the lenghts of orbits, not even for bi-quadratic cooperative systems.

\subsection{Proof of part~(i)}

 The proof uses a construction similar to a small Turing machine operating on several long circular tapes.
 We will first introduce the main idea of the construction, but without requiring
 the system to be cooperative and bi-quadratic.
 Subsequently we will show how to modify the construction  so that the network will also be cooperative, bi-quadratic and will have a strongly connected
 digraph.

\subsubsection{A Simple Counting Model}  \label{section simple}

In this subsection we consider a conceptual model of a (not necessarily bi-quadratic or cooperative) Boolean network with
orbits of length $2^N$, for arbitrary $N>0$.  We also discuss the problems that are involved in constructing such a network under the restrictions of Theorem~\ref{theorem counterexample}(i).  Consider a Boolean system with states $(s_1,\ldots , s_N)$ and the dynamics defined by

\begin{equation} \label{simple cycle}
\begin{array}{l}
s_i(t):=s_{i+1}(t-1), \ \ \  i=1,\ldots, N-1,  \\
s_N(t):=\gamma(s_1(t-1),mode(t-1)).
\end{array}
\end{equation}

One can think of $\gamma$
 as implemented by a Turing machine operating on variables numbered $i = 1, \ldots , N$ whose values are written on a
circular tape.
The variable $mode$ can have one of two possible values for every $t$, namely $mode=\mbox{\it{rotate}}$, and $mode=\mbox{\it{switch}}$, and the function~$\gamma$ is defined by

\begin{equation} \label{simple function}
\begin{array}{l}
\gamma(x,\mbox{\it{rotate}})=x,   \\
\gamma(x,\mbox{\it{switch}})=1-x.
\end{array}
\end{equation}

Thus while $mode(t)=\mbox{\it{rotate}}$, iterating this machine will cyclically rotate the values of $s_1,\ldots , s_N$.  Whenever $mode=\mbox{\it{switch}}$, the machine also will rotate the variable values, but it will invert them at the site $s_N$.

Now let us define the value of the variable $mode$, in such a way that this machine behaves like a counter in base two.  Let us require that
$mode(t)=\mbox{\it{switch}}$ at the times $t=0,N,2N,3N,\dots$. A possible mechanism for ensuring this property will be discussed when we present our modified construction. For all other times $t$, define

\begin{equation} \label{simple mode}
mode(t):=\left\{ \begin{array}{ll}
mode(t-1), & \mbox{ if } s_1(t-1)=1, \\
\mbox{\it{rotate}}, &  \mbox{ if } s_1(t-1)=0.
\end{array} \right.
\end{equation}

Thus the model turns into \mbox{\it{switch}} mode exactly at the times $t=0,N,2N,\ldots$, and it only returns back to \mbox{\it{rotate}} mode after $s_1(t_1)=0$ for some $t_1>t$.  The following lemma shows in what way this machine is a counter:  if the states of the system encode numbers in binary format appropriately, then $N$ iterations are equivalent to the addition of one unit modulo $2^N$.

\begin{lemma} \label{lemma simple count}
Given any state $s$ of the model, define $\alpha(s):=s_1 2^0 + s_2 2^1 + \ldots + s_N 2^{N-1}$.  Then $\alpha(s(N))=\alpha(s(0))+1$ mod $2^N$.
\end{lemma}

\noindent
\textbf{Proof:}
Consider an initial state $s(0)$ and let $j\geq 0$ be such that $s_i(\eta)=1$, for $1\leq \eta\leq j <N$, and $s_{j+1}(0)=0$.   Note that $\alpha(s(0))<2^N-1$ in this case.
We have $mode(0)=\mbox{\it{switch}}$ by the definition above (\ref{simple mode}).  By (\ref{simple cycle}), $s_1(\eta)=1$ for $0\leq \eta\leq j-1$ and $s_1(j)=0$.  Therefore $mode(\eta)=\mbox{\it{switch}}$, for $1\leq \eta\leq j$, and $mode(j+1)=\ldots=mode(N-1)=\mbox{\it{rotate}}$.   At time $t=N$, the variable values have completed a full rotation and returned to their starting points, except that $s_\eta=0$ for $1\leq \eta\leq j$, $s_{j+1}=1$, and $s_{j+2},\ldots , s_N$ are unchanged.  Clearly $\alpha(s(N))=\alpha(s(0))+1$ in this case.

It remains to show the result for the case $j=N$, i.e.\  $s_i(0)=1$, for every $i=1,\ldots,N$.  In that case $mode(0)=mode(1)=\ldots = mode(N-1)=\mbox{\it{switch}}$ by (\ref{simple cycle}) and (\ref{simple mode}).  In this way every value of the system is inverted at $s_1$ from 1 to 0, so that $s_i(N)=0$ for $i=1\ldots N$.  Therefore $\alpha(s(N))=0=\alpha(0)+1$ mod $2^N$. $\Box$
\bigskip

\begin{corollary} \label{corollary simple count}
 The network given by equations (\ref{simple cycle}), (\ref{simple function}), (\ref{simple mode}), contains an orbit of length at least $2^N$.
\end{corollary}
\noindent
\textbf{Proof:}  Since the variable $mode$ is reset to $\mbox{\it{switch}}$ for $t=0,N,2N,\ldots$,  Lemma~\ref{lemma simple count} applies
 at each of these time points.  Therefore one can start with $s(0)=0$, and apply
Lemma~\ref{lemma simple count} successively to reach states $s(0),s(N),s(2N), \ldots,$ $s((2^N-1)N)$, which are all different from each other.
$\Box$
\bigskip

Importantly, the function $\gamma$ negates the values of the input $x$ in switching mode.   This appears to be an essential non-cooperative component (or negative feedback) of this system.  Nevertheless, it is shown below that in fact one can rewrite our system in such a way that the resulting system is cooperative.

\subsubsection{A Generalized Counter}

Before proceeding with the proof of the main result, consider the following generalization of the simple counter above.   Instead of individual Boolean values, each variable $s_i$ is now considered to be a vector with $\ell>1$ Boolean entries, $S_i=(s_i^\ell,\ldots,s_i^1)$.
We will treat $S_i$ as a binary code for a nonnegative integer $< 2^\ell$. At each time $t$, the system continues to be in one of two modes $mode(t)=\mbox{\it{switch}}$ or $mode(t)=\mbox{\it{rotate}}$, but the function $\gamma$ is now replaced with a vector function $G$ which we describe in the next paragraph.

As before, when $mode=\mbox{\it{rotate}}$ we let $G(x,mode):=x$. When $mode=\mbox{\it{switch}}$,
and given $S=(s^\ell,s^{\ell -1}\ldots,s^1)\not=(1,\ldots,1)$, let $j$ be such that $s^{\eta}=1$ for $1\leq \eta \leq j<l$ and $s^{j+1}=0$.  Define $R$ by letting $r^\eta:=0$    for
$1\leq \eta \leq j$, letting $r^{j+1}:=1$, and $r^\eta:=s^\eta$ for $j+1<\eta\leq \ell$.  Set $G(S,\mbox{\it{switch}}):=R$. If $S=(1,\ldots,1)$, set $G(S,\mbox{\it{switch}}):=(0,\ldots,0)$.   In other words, the function $G(S,\mbox{\it{switch}})$ is defined as the addition of~1 to the vector $S$, in base 2 and modulo~$2^l$.

We define the generalized system

\begin{equation} \label{vector cycle}
\begin{array}{l}
S_i(t):=S_{i+1}(t-1), \ \ \  i=1,\ldots, N-1,  \\
S_N(t):=G(S_1(t-1),mode(t-1)),
\end{array}
\end{equation}
where $G$ is defined as above.  The variable $mode(t)$ has the value $\mbox{\it{switch}}$ for $t=0,N,2N,\ldots$ and for other values of $t$:

\begin{equation} \label{vector mode}
mode(t):=\left\{ \begin{array}{ll}
mode(t-1), & \mbox{ if } S_1(t-1)=(1,\ldots,1), \\
\mbox{\it{rotate}}, &  \mbox{ otherwise. }
\end{array} \right.
\end{equation}

One can naturally think of this system as a Turing machine that computes $G$ and operates on $\ell$ simultaneously advancing circular tapes, with
$S_i$ representing the $i$-th cross-section of these tapes.  The machine reads the value of $S_1$ and writes to $S_N$.

\begin{lemma} \label{lemma vector}  The network defined by equations (\ref{vector cycle}), (\ref{vector mode}) contains an orbit of length at least $2^{N \ell}$.
\end{lemma}

\noindent
\textbf{Proof:}
For $S = (s^\ell,\ldots, s^1)\in \{0,1\}^l$, define $\beta(S):=s^1 2^0  + s^2 2^1 +\ldots + s^\ell 2^{\ell-1}$.  Note that $\beta(G(x,\mbox{\it{switch}}))=\beta(x)+1$ mod $2^\ell$.   We follow an argument very analogous to Lemma~\ref{lemma simple count} and Corollary~\ref{corollary simple count}.  Let $\alpha(S):=\beta(S_1) (2^\ell)^0 + \beta(S_2) (2^\ell)^1 + \ldots + \beta(S_N) (2^\ell)^{N-1}$.  Thus the vector $(\beta(S_1),\ldots,\beta(S_N))$ can be regarded as the representation of $\alpha(s)$ in base $2^l$.

As in the proof of Lemma~\ref{lemma simple count}, consider an initial state $S(0)$, and let $j\geq 0$ be such that $S_\eta(0)=(1,\ldots,1)$, for $1\leq \eta\leq j <N$, and $S_{j+1}(0)\not=(1,\ldots,1)$.   As before, we have $mode(\eta)=\mbox{\it{switch}}$ for $0\leq \eta\leq j$, and
 $mode(j+1)=\ldots=mode(N-1)=\mbox{\it{rotate}}$.   At time $t=N$ we have $S_\eta=(0,\ldots,0)$ for $1\leq \eta\leq j$,
as well as $\beta(S_{j+1})=\beta(S_{j+1}(0))+1$, and $S_{j+2},\ldots , S_N$
 are unchanged from $t=0$.  Clearly $\alpha(S(N))=\alpha(S(0))+1$.

In the case that $S_i(0)=(1,\ldots,1)$ for every $i=1,\ldots,N$, it follows as before that $mode(0)=mode(1)=\ldots  = mode(N-1)=\mbox{\it{switch}}$.  Therefore $S_i(N)=(0,\ldots,0)$ for $i=1\ldots N$, and $\alpha(S(N))=0$.

Repeating this process for $S(0)\equiv 0$ and $t=N,2N,\ldots,$ as in Corollary~\ref{corollary simple count}, one finds states $S$ of the system such that $\alpha(S)=1,2,\ldots$, and which are therefore pairwise different.
When $S_i=(1,\ldots,1)$ for all $i$, that is, when $\alpha(S(t))=(2^\ell)^N-1$, this process reverts to $\alpha(S(t+N)) = 0$. $\Box$
\bigskip

\subsubsection{A Cooperative Counter}

In this subsection we carry out a construction which is analogous to that in Subsection~\ref{section simple}, but in which the underlying Boolean network is cooperative, bi-quadratic, and has a strongly connected digraph.  We will need to define some auxiliary Boolean networks with designated input and output variables.

Throughout this section let $L>0$ be an arbitrary even number, and consider the set $A:=\{(a_1,\ldots a_L)\in \{0,1\}^L\,|\, a_1+\ldots+a_L=L/2\}$.   Define the special sequences $\mbox{START}=(1,\ldots,1,0,\ldots,0)$, i.e., $L/2$ ones followed by $L/2$ zeros, and similarly $\mbox{ACTIVE}=$ $(0,\ldots,0,1,\ldots,1)$.  The idea of the proof is to code arbitrary binary vectors of length $\ell$ as elements of $A$.  Similarly, the internal variable $mode$ will be encoded by a Boolean vector $d$ of length~2, with $(1,0)$ standing for $switch$ and $(0,1)$ standing for $rotate$.  This will allow us to implement the dynamics described in the previous subsection in a cooperative system.

\begin{lemma}  \label{lemma transform}
Let $g:A\to A$ be an arbitrary function.  There exists a Boolean network $B$ with input vectors $a=(a_1,\ldots, a_L)$, $d=(d_1,d_2)$, and output vector $c=(c_1,\ldots, c_L)$, such that for some fixed $m>0$ the following equation holds for every $t$ and $a(t)\in A$, regardless of the initial state of $B$:

\begin{equation} \label{eq B module}
c(t+m):=\left\{ \begin{array}{ll}
a(t),    &  \mbox{ if } d(t)=(0,1),  \\
g(a(t)), &  \mbox{ if } d(t)=(1,0).
\end{array} \right.
\end{equation}

Furthermore, the network $B$ is cooperative, every node of its associated digraph has in- and outdegree of at most 2, and the indegree (outdegree) of every designated input (output) variable is 0.
\end{lemma}

\noindent
\textbf{Proof:}
Define the set $\hat{A}:=A\times\{ (0,1),(1,0)\}$, and the function $\hat{h}:\hat{A}\to A$ by $\hat{h}(a,(1,0)):=g(a)$, $\hat{h}(a,(0,1)):=a$, for arbitrary $a\in A$.  Since $\hat{A}$ is an unordered set in the coordinatewise partial order of Boolean vectors,
$\hat{h}$ can be extended to a cooperative function $h:\{0,1\}^{L+2}\to \{0,1\}^L$ \cite{Just:Enciso:embedding}.   The result will follow from building a suitable Boolean network that computes the function $h$.

Consider a fixed component $h_i:\{0,1\}^{L+2}\to \{0,1\}$ of $h$.  By the cooperativity of this function, one can write it in the normal form $h_i(y_1,\ldots,y_{L+2})=\Psi_1^i(y_1,\ldots,y_{L+2})\vee\ldots \vee \Psi_{k_i}^i(y_1,\ldots,y_{L+2})$, where each $\Psi_j^i$ is the conjunction of a number of variables, i.e., $\Psi_j^i(y_1,\ldots,y_{L+2})=y_{\alpha_{1i}}\wedge\ldots \wedge y_{\alpha_{ji}}$.
This suggests a way of computing $h_i$: define Boolean variables $\psi_j^i(t):=\Psi_j^i(y(t-1))$, and then let $h_i(t):=\psi_1^i(t-1) \vee\ldots \vee \psi_{k_i}^i(t-1)$.   Repeating this procedure for all components of $h$ yields a Boolean network which computes $h$ in $m=2$ steps, and which is cooperative and has indegree (outdegree) 0 for every input (output).

In order to satisfy the condition that every node have in- and outdegree of at most 2, we need to modify this construction by introducing additional variables.  First, note that the outdegree of every input $y_i$ can be very large.  One can define two additional variables which simply copy the value of $y_i(t)$, then four variables that copy the value of the previous two, etc.  This procedure is repeated for each $y_i$ so that at least as many copies of each variable are present as appear in the expressions of all $\psi_j^i$.  A similar cascade can be used to define each $\psi_j^i$ and  $h_i$ so that each indegree is at most two.  If $\psi_i^j=y_{\alpha_1}\wedge y_{\alpha_2} \wedge y_{\alpha_3}$, say, then one can define $z_1(t):=y_{\alpha_1}(t-1)$, $z_2(t):=y_{\alpha_2}(t-1)\wedge y_{\alpha_3}(t-1)$, $\psi_i^j(t):=z_1(t-1)\wedge z_2(t-1)$.
Similarly for longer disjunctions and each $\psi_j^i$ and also similarly for $h_i$, in which case $\wedge$ is replaced by $\vee$ at each step.  This produces a computation of $h_i$ in $m_i$ steps for each $i$.  Finally, after introducing further additional variables at each component $i$ if necessary to compensate for unequal lengths of the expressions for $\psi^i_j$,
the Boolean vector $h(y_1,\ldots, y_{L+2})$ can be computed in exactly $m=\max(m_1,\ldots,m_{L})$ steps.
$\Box$
\bigskip

\begin{remark}\label{remus}
Without loss of generality, we can assume that for every state variable $s$ in the network $B$, there exists some input variable $d_i$ or $a_i$ and a directed path from this input towards $s$. Similarly, we can assume that for every state variable $s$, there exists an output variable~$c_i$ such that there is a directed path from $s$ to $c_i$.
\end{remark}

If that wasn't the case for some $s$, one could delete $s$ from the system without altering equation (\ref{eq B module}).   By choosing a suitable coding of integers $< 2^\ell$ we may assume that the Boolean function $g$ to which we will apply Lemma~\ref{lemma transform} is such that

\begin{equation}\label{gprop}
g(1,0,\ldots,1,0)=(0,1,0,1,\ldots,0,1).
\end{equation}

It follows that each $h_i$ as in the proof of Lemma~\ref{lemma transform} is non-constant and no output variable will be deleted.

Lemma~\ref{lemma transform}
 can be used to compute a function $g^*$ which will be used in a way analogous to $\gamma$ in equation (\ref{simple cycle}).
More precisely, let $\Gamma$ be an injection that maps the set of integers $\{0, \ldots , 2^\ell - 1\}$ into $A \backslash \{\mbox{START}\}$ in such a way that $\Gamma(2^\ell-1) = \mbox{ACTIVE}$.  Such a function exists as long as $L$ is sufficiently large relative to $\ell$; we will describe a suitable choice for~$L$ in the next subsection.  Let $g: A \rightarrow A$ be such that $g(a) = \Gamma(\Gamma^{-1}(a) \oplus 1)$, where $\oplus$ denotes addition modulo $2^\ell$,
 for $a$ in the range of $\Gamma$, and $g(\mbox{START}) = \mbox{START}$.  Let $g^*$ be the corresponding function on
$A \times \{(0,1),(1,0)\}$
given by Lemma~\ref{lemma transform}.

Now define $S_i = (a^1_i, \ldots , a^L_i)$ for $i \in [N]$ and let

\begin{equation} \label{r cycle}
\begin{array}{l}
S_i(t):=S_{i+1}(t-1), \ \ \  i=1,\ldots, N-1,  \\
S_N(t):=g^*(S_1(t-1),d(t-1)),
\end{array}
\end{equation}

Define the dynamics for the variable $mode(t) = d(t)$ by

\begin{equation} \label{r mode}
d(t):=\left\{ \begin{array}{ll}
(1,0), & \mbox{ if } S_1(t-1)=\mbox{START}, \\
d(t-1), & \mbox{ if } S_1(t-1)=\mbox{ACTIVE}, \\
(0,1), &  \mbox{ otherwise. }
\end{array} \right.
\end{equation}

Again, one can naturally think of this system as a Turing machine that operates on $L$ simultaneously advancing circular tapes, with
$S_i$ representing the $i$-th cross-section of these tapes.  The machine reads the value of $S_1$ and writes to $S_N$.

If this system starts in a state where $S_1 = \mbox{START}$ and $S_i$ is in the range of $\Gamma$ for all $i > 1$, then the dynamics on these
tapes will code the dynamics of the system described in the previous subsection.  In particular, Lemma~\ref{lemma vector} implies that such states
are contained in orbits of length at least $2^{\ell(N-1)}$.

It remains to show that the dynamics on $d$ described by~(\ref{r mode}) can be implemented in such a way that the whole system becomes bi-quadratic, cooperative, and has a strongly connected digraph.
Unfortunately,  Lemma~\ref{lemma transform} cannot be used for this purpose because the desired output depends not only on $S_1(t-1)$  but on the history (of unknown length) of~$S_1$ since the last time when $S_1$ took the value $START$.  This history is summarized by the value of $d(t-1)$, but the problem is that~(\ref{r mode}) has
$n$ inputs, with $d(t-1)$ acting as input for the computation of $d(t)$, which poses a problem
for implementation by quadratic functions.  The following lemma shows how this problem can be solved.

\begin{lemma} \label{lemma switch}
There exists $\mu>0$ and a Boolean network $D$ with input vector $p=(p_1,\ldots, p_L)$, and output vector $q=(q_1,q_2)$, such that the following holds for any initial condition of $D$.   Consider any sequence of inputs $p(0),p(1), \ldots, p(M)$, $M>1$, such that

\noindent
i)  $p(t)\in A$, for $0\leq t\leq M$,

\noindent
ii) $p(0)=\mbox{START}$,
and

\noindent
iii) $p(t)\not=\mbox{START}$, for $0<t\leq M$.

\noindent
Let $j\geq 0$ be such that $p(t)=\mbox{ACTIVE}$ for $1\leq t\leq j$,  $p(j+1)\not=\mbox{ACTIVE}$ (or $p(1)=\ldots =p(M)=\mbox{ACTIVE}$ and $j=M$).
Then
\begin{equation}  \label{eq switch}
q(t)=\left\{ \begin{array}{ll}
(1,0), &  \mu\leq t \leq \mu+j,  \\
(0,1), &  \mu+j < t \leq  \mu+M,
\end{array}  \right.
\end{equation}

Furthermore, the network $B$ is cooperative, every node of its associated digraph has in- and outdegree of at most 2, and the indegree (outdegree) of every designated input (output) variable is 0.
\end{lemma}

\begin{figure}[ht]
\centerline{\includegraphics[width=4.5in]{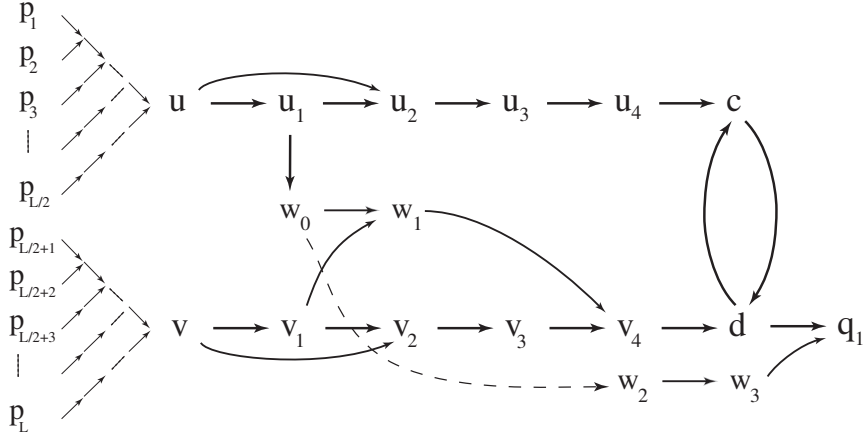} }
\caption{The digraph of the network $D$ which is used to compute the output $q_1$ from the input $p$.  The formulas for each interaction (i.e. $\wedge,\vee$)  are omitted in this figure.}
\label{figure switch}
\end{figure}

\noindent
\textbf{Proof:}
The idea of this proof is based on the simple system $c(t)= u(t-1) \vee d(t-1)$, $d(t)= v(t-1) \wedge c(t-1)$, with inputs $u,v$.  This switch $d$ is turned on by letting both inputs $u=1$ and $v=1$ for a short time, after which $u$ can be turned to $0$ while $v$ is left equal to~$1$.  After letting $v=0$ for a short time, the switch resets and doesn't restart even if $v=1$ again.

Let $t=0$ without loss of generality; the more general case being completely analogous.   For the sake of clarity assume for now that $0<j<M$,
but the same construction allows for $j=0$ and $j=M$ as described below.  See Figure~\ref{figure switch} which displays the circuit described below.
Define for the moment $u(t):=p_1(t-1) \wedge \ldots \wedge p_{L/2}(t-1)$,
$v(t):=p_{L/2+1}(t-1)\wedge \ldots \wedge p_{L}(t-1)$ (a modification of this definition with additional variables and indegree two is displayed in the
figure and described below).  Thus $u(t)=1$ if and only if $p(t-1)=\mbox{START}$, and $v(t)=1$ if and only if $p(t-1)=\mbox{ACTIVE}$, since by assumption $p(t)\in A$.

Define
\[
u_1(t):=u(t-1),\ u_2(t):=u(t-1)\vee u_1(t-1), \ u_3(t):=u_2(t-1),\ u_4(t):=u_3(t-1),
\]
\[
v_1(t)=v(t-1), \ v_2(t)=v(t-1)\wedge v_1(t-1),\ v_3(t):=v_2(t-1),\ v_4(t):=v_3(t-1)\vee w_1(t-1),
\]
\[
w_0(t):=u_1(t-1),\ w_1(t):=w_0(t-1)\wedge v_1(t-1),
\]
\[
c(t):=u_4(t-1)\vee d(t-1), \ d(t):=v_4(t-1) \wedge c(t-1).
\]

Intuitively, $u_4$ is a time-transposed copy of $u$ where every 1 has been doubled due to the feed-forward loop at $u_2$.  Also, $v_4$ is similar to a time-transposed copy of $v$ where every 0 has been doubled. The auxiliary variables $w_i$ only play a role at a single time step as described below.  The loop $c \leftrightarrow d$ forms the core of the switch in the system.

A simple calculation shows that $u_4(4)=u_4(5)=1$, $u_4(t)=0$ for $5<t\leq  M+4$.  On the other hand, since $v(1)=0, v(2)=\ldots=v(1+j)=1, v(2+j)=0$, we infer that
$v_2(2)=v_2(3)=0$, $v_2(t)=1$ for $3<t\leq 2+j$, $v_2(3+j)= v_2(4+j) = v_3(4+j)= v_3(5+j) = 0$.   It follows that
$w_1(3)=0$ (since $v_1(2)=0$),
 and that $w_1(4)=1$ if and only if $v_1(3)=1$ (since $w_0(3)=1$).    This in turn holds since we are assuming for now that $j>0$.  Also, $w_1(s)=0$ for $s>4$.

We use the data for $w_1$ and $v_3$ to compute the values of $v_4$.  From $w_1(3)=v_3(3)=0$, it follows that $v_4(4)=0$.  From $w_1(4)=1$ it follows that $v_4(5)=1$, and using $v_3$ we similarly infer that $v_4(t)=1$ for $4< t\leq 4+j$.  Also, $v_4(5+j)=v_4(6+j)=0$.

We conclude that $c(5)=1$, $d(5)=0$, regardless of the  values of $c,d$ at earlier time steps.
  Since $j>0$,
one has $c(6)=1$, $d(6)=1$, and in general $c(t)=d(t)=1$ for $5<t\leq 5+j$.  Then $c(6+j)=1$, $d(6+j)=0$ (because $v_4(5+j) = 0$).  It follows that $c(t)=d(t)=0$ for $7+j\leq t\leq 5+M$, and $d(6+M)=0$.

In particular  $d(t)=1$ for exactly $j$ time steps, $5<t\leq 5+j$, and then  $d(t)=0$ for $6+j\leq t \leq 6+M$.
Since we want the variable $q_1$ to be equal to 1 during exactly $j+1$ time steps, we define the additional variables
\[
w_2(t):=w_0(t-1),\ w_3(t):=w_2(t-1),\ q_1(t):=w_3(t-1)\vee d(t-1).
\]
Calculating that $w_3(5)=1$, $w_3(t)=0$ for $5<t\leq 5+M$, we conclude that $q_1(t)=1$ for $6\leq  t\leq 6+j$, and $q_1(t)=0$ for $6+j<s\leq 7+M$.

The case $j=0$ is very similar to the one above, except that $w_1(4)=0$ (instead of 1 for $j>0$), $v_4(4)=v_4(5)=0$, and therefore $d(t)=0$ on all $6\leq t\leq M+6$.  Thus $q_1(6)=1$, and $q_1(t)=0$ for larger values of $t$.

 In the case $j=M$, one can compute $v_4(t)=1$ for $5\leq t< M+5$.  This allows the variables $c(t),d(t)$ to remain equal to 1 up to and including $t=M+5$.  Therefore $q_1(1)=1$ up to and including $t=6+M$.

In order to define the variable $q_2$, it suffices to use a construction dual to the previous one (recall that simply negating $q_1$ is not permitted).  That is, define $\hat{u}(t):=p_{L/2+1}(t-1)\vee \ldots \vee p_{L}(t-1)$, and $\hat{v}(t):=p_1(t-1)\vee \ldots \vee p_{L/2}(t-1)$, in such a way that  $\hat{u}(t)=0$ if and only if $p(t-1)=\mbox{START}$, and $\hat{v}(t)=0$ if and only if $p(t-1)=\mbox{ACTIVE}$.   Define variables $\hat{u}_1,\hat{v}_1$ etc.\  similarly as above, except that every $\wedge$ in the function definition is replaced by $\vee$ and vice versa.
Then it will necessarily follow that $q_2= \neg q_1$
 on the interval $6\leq t\leq 6+M$.   Using the value $\mu=6$, equation (\ref{eq switch}) is satisfied.

Notice that the system described so far is cooperative, and that all in- and outdegree requirements are satisfied except for the indegree of the variables $u,v,\hat{u},\hat{v}$.   These terms can now be replaced in a routine manner by a cascade of variables (see Figure~\ref{figure switch}),  in such a way that $u(t)=1$ if and only if $p(t-\tau)=\mbox{START}$, etc., for some $\tau>1$.  This will increase the delay $\mu$ but leave the computations and the other properties of this system unchanged.
$\Box$
\bigskip

Using the function $g$ defined above, we consider the cooperative networks $B$ and $D$ from Lemmas~\ref{lemma transform} and~\ref{lemma switch}.
Recall that $B$ ($D$) has variables $a,d$ ($p$) which are specifically designated as inputs, variables $c$ ($q$) specifically designated as outputs, and a  `processing delay' $m$ ($\mu$).  The cooperative network, which will be denoted by $\cS$, is defined by  $B$ and $D$, together with the equations

\begin{equation} \label{cooperative cycle}
\begin{array}{l}
S_i(t):=S_{i+1}(t-1), \ \ \  i=m+2,m+3,\ldots, N,  \\
S_{N+1}(t):=c(t-1),
\end{array}
\end{equation}
and

\begin{equation} \label{cooperative Turing}
\begin{array}{l}
a(t):=S_{m+2}(t-1),  \\
d(t):=q(t-1),        \\
p(t):=S_{m+\mu+2}(t).
\end{array}
\end{equation}

\begin{figure}[ht]
\centerline{\includegraphics[width=6in]{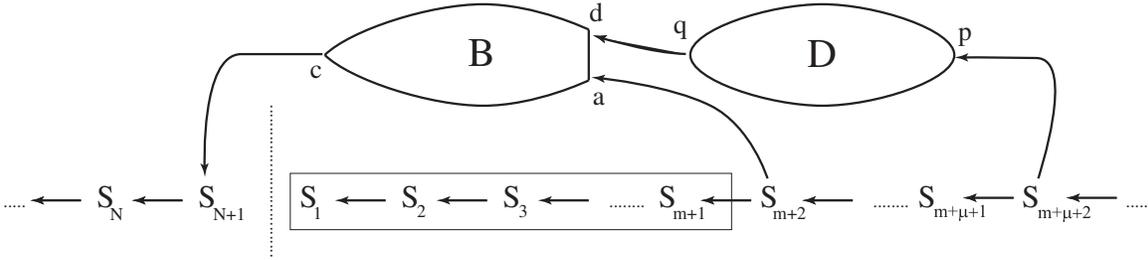} }
\caption{The network interconnections of the system $\cS$ given by $B$, $D$, and equations (\ref{cooperative cycle}), (\ref{cooperative Turing}).  The variables $S_1,\ldots, S_{m+1}$ are displayed in a box to indicate that they are not part of $\cS$ but only included in the proof of Lemma~\ref{teo cooperative cycle}.}
\label{figure cooperative}
\end{figure}

See Figure~\ref{figure cooperative} for an illustration.
In order to get an orbit of length $2^{\ell N}$, we need to code integer $\{0, \ldots , N \ell - 1\}$ in $N$ blocks of the form $S_i$, and we need an additional block $S_j$ to contain START and thus mark the beginning of the coded sequence.
Since both of the subnetworks used in the construction of this system contain only the Boolean operators $\wedge,\vee$ in their expression (and no negations), it follows from (\ref{cooperative cycle}) and (\ref{cooperative Turing}) that the same is the case for the full network, hence the system is cooperative.

\begin{proposition}   \label{teo cooperative properties}
The digraph of the Boolean network $\cS$ is strongly connected and bi-quadratic.
\end{proposition}

\noindent
\textbf{Proof:}
The fact that every in- and outdegree is at most 2
 follows directly from equations (\ref{cooperative cycle}), (\ref{cooperative Turing}) and Lemmas~\ref{lemma transform} and~\ref{lemma switch}, taking into account that the indegree (outdegree) of every input (output) variable is zero within their respective subnetwork.
See also Figure~\ref{figure cooperative}.

In order to show the strong connectivity of the digraph, first we show that there exists a directed path from every node in the network to the node $q_1$, the first component in the output of $D$.  It is clear from the circuit defining $D$ that every input variable $p_i$ has a path connecting to $q_1$ (the first $L/2$ components of $p$ through the variables $u,u_1,\ldots$ and the last $L/2$ components through $v,v_1,\ldots$).  Therefore every variable in $S_i$ can reach $q_1$ as well.  By Remark~\ref{remus}, the same applies to every variable of $c$, and thus to every variable in the subnetwork $B$.
Thus the same applies also to $q_2$, and hence to every state in the subnetwork $D$.

Now we show that there exists a path from $q_1$ to every node in the network.  Suppose first that there exists $c_j$ such that neither $d_1$ or $d_2$ contains a path towards $c_j$.
This would imply that $g_j(x)=x_j$ for every argument $x\in A$, by equation (\ref{eq B module}).  But this is not possible if $g$ is chosen so that~(\ref{gprop}) holds.  Thus for every $j$, there exists a path from either $d_1$ or $d_2$ to $c_j$ (and therefore from $q_1$ or $q_2$ to $c_j$).

Since there exists a path from $q_1$ to $q_2$, it follows that there is a path from $q_1$ to every~$c_j$.  Thus every component of every state $S_i$, $p$, and $a$ can be reached by a path from $q_1$.  Every state in $B$ can be reached from $d_1$ and hence $q_1$, once again by Remark~\ref{remus}; the same applies to $q_2$, and every state in the subnetwork $D$.
$\Box$

\begin{lemma}  \label{teo cooperative cycle}
Let $N \geq 3m + 3\mu + 5$. Then the system $\cS$ has an orbit of length greater than or equal to $2^{\ell N}$.
\end{lemma}

\noindent
\textbf{Proof:}
Let $\cS^+$ be the Boolean network obtained from $\cS$ by adding blocks of variables $S_1, \ldots , S_{m+1}$ of size $L$ each (as shown in
Figure~\ref{figure cooperative}), with $S_i(t):=S_{i+1}(t-1)$ as in~(\ref{cooperative cycle}) also holding for $i = 1, \ldots , m+1$.  These variables cannot change the length of the original system's orbits
since they don't feed back into it, but they can nevertheless be used for the study of the network.
 Let us call a state $S^+(0)$ of $\cS^+$  \emph{pre-canonical} if $S_{m+\mu+2}^+(0) = \mbox{START}$ and $S_i^+(0) \in range(\Gamma)$
for $i \neq m+\mu+2$. Let us call a state $S^*(0)$ of $\cS^+$  \emph{canonical} if there exists a pre-canonical state $S^+(0)$ so that $S^*_j(0) \in range(\Gamma)$ for all $j$ with $N-m - \mu-1 \leq j \leq N+1$ and $S^*(0)$ and $S^+(m+\mu+1)$ agree on all remaining nodes.  Note that our assumption on $N$ implies in particular that $S^*_i(0) = S^+_i(n+\mu+1)$ for all $i = 1, \ldots 2m + 2\mu + 4$.
A state of $\cS$ that can be obtained by removing $S_1 \cup \dots \cup S_{m+1}$ from a canonical state of $\cS^+$ will be called a \emph{proper state.}

We will show that every canonical state of $\cS^+$ belongs to an orbit of length at least $2^{\ell N}$ of $\cS^+$.  Since the variables in
$S_1 \cup \dots \cup S_{m+1}$ do not act as inputs to variables in $\cS$, it will follow that every proper state of of $\cS$ belongs to an orbit of length at least $2^{\ell N}$ of $\cS$.

So let $S^*(0)$ be a canonical state of $\cS^+$, and let $S^+(0)$ be a corresponding pre-canonical state.  After $\mu$ iterations we will have
$S^+_{m+2}(\mu) = S^+_{m+\mu+2}(0) = \mbox{START}$ and $q^+(\mu) = (1,0)$ by the choice of subnetwork $D$ and Lemma~\ref{lemma switch}.  Thus
$d^+(\mu+1) = (1,0)$ and $a^+(\mu+1) = \mbox{START}$ by (\ref{cooperative Turing}).  By the choice of subnetwork $B$ and Lemma~\ref{lemma transform} and the assumed relationship between $S^*$ and $S^+$ we will have $c^*(0) = c^+(m + \mu + 1) = S^+_1(m + \mu + 1) = \mbox{START} = S^*_1(0)$.  By~(\ref{cooperative cycle}), this implies $S^*_{N+1}(1) = \mbox{START}$.

More generally, let us define for $t \geq 1$ the value of $mode(t)$ as $q^+(t + \mu)$ if $t \leq m + \mu$ and as
$q^*(t - m - 1)$ if $t \geq m + \mu + 1$.  Let $j$ be such that $S^*_i(1) = \mbox{ACTIVE}$ for $1 \leq i \leq j$ and $S^*_{j+1} \neq \mbox{ACTIVE}$.  It follows from the choice of subnetwork $D$ and Lemma~\ref{lemma switch} that

\begin{equation} \label{eq mode cooperative}
mode(t)=(1,0) = \mbox{\it{switch}},\ \ 0\leq t\leq j;\ \ mode(t)=(0,1) = \mbox{\it{rotate}},\ \ j+1\leq t\leq N.
\end{equation}

Similarly, by the choice of subnetwork $B$ and Lemma~\ref{lemma transform}

\begin{equation} \label{cooperative mode}
S^*_{N+1}(t):=\left\{ \begin{array}{ll}
S^*_1(t-1), & \mbox{ if } mode(t-1)=\mbox{\it{rotate}}, \\
g(S^*_1(t-1)), &  \mbox{ if } mode(t-1)=\mbox{\it{switch}},
\end{array} \right.
\end{equation}

Thus when starting in a canonical initial state, our system behaves exactly as specified in~(\ref{r cycle}) and~(\ref{r mode}) and the lemma follows.
$\Box$
\bigskip

\subsubsection{The Choice of $\ell$ and $L$}

We can use Lemmas~\ref{teo cooperative properties} and~\ref{teo cooperative cycle} to prove the theorem stated in the introduction.  Let $0< c<2$ be arbitrary.  We prove first that there exist $L>0$ even and an integer $\ell>0$ such that

\begin{equation} \label{L l condition}
\binom{L}{L/2} > 2^\ell > c^L.
\end{equation}

The second inequality is equivalent to $L/\ell < \ln 2/ \ln c$; thus let $L$ be an even integer with $L=w \ell$, for some fixed $1<w<\ln 2/ \ln c$. Using Stirling's formula, we have $\binom{L}{L/2} > v\, 2^L/\sqrt{2\pi L}$ for large enough $L$, where $0<v<1$ is arbitrary and fixed.  The first inequality in (\ref{L l condition}) is satisfied if $v\, 2^L/\sqrt{2\pi L}>2^\ell$. But after replacing $L=w \ell$ this is equivalent to $2^{(w-1)\ell}> v^{-1}\sqrt{2\pi w \ell}$.  Clearly this inequality is satisfied for sufficiently large $\ell$, hence (\ref{L l condition}) follows.

The first inequality is now used to carry out the construction of system $S$, which by Lemmas~\ref{teo cooperative properties} and~\ref{teo cooperative cycle} is cooperative and bi-quadratic with strongly connected digraph, and has an
orbit of length greater than or equal to $2^{\ell N}$.
 It remains to show that $2^{\ell N } \geq c^n$ for sufficiently large $N>0$, where $n$ is the dimension of the system.

Let $T$ be the total
number of variables in the subnetworks $D,B$.  Note that $T$ depends only on $L,\ell$, and not on $N$.  Then $n=(N+1 - (m+1)L+T=NL-mL+T$.  Notice that $c^n\leq 2^{N \ell}$ if and only if $(NL-mL+T)\ln c \leq N l \ln 2$, which holds if and only if

\[
L \ln c \leq \ell \ln 2 + \frac{mL-T}{N} \ln c.
\]
But this equation is satisfied for large enough $N$, since $L \ln c< \ell \ln 2$ by (\ref{L l condition}). $\Box$
\bigskip

\subsection{Proofs of parts~(ii) and~(iii)}

Let $(\Sigma, f)$ be a bi-quadratic cooperative Boolean system  of dimension $n-2$ that contains an orbit of length $c^n$.  Let $\Pi = \{0, 1\}^{[n]}$, let $g_k = f_k$ whenever $k < n-1$ and $f_k$ is strictly quadratic, let
$g_k = s_{i_k} \wedge s_n$ whenever $k < n$ and $f_k = s_{i_k}$, and let $g_{n-1} = g_n = s_{n-1} \wedge s_n$.  Then $(\Pi, g)$ is cooperative, quadratic, and has only strictly quadratic regulatory functions.
Now let $s \in \Sigma$
be a state in an orbit of length at least $c^n$ of $(\Sigma , f)$, and define a state  $s^* \in \Pi$ by
$s^* = [s_1, \ldots, s_{n-2}, 1, 1]$.  Then the orbit of $s^*$ in $(\Pi, g)$ has the same length as the orbit of
$s$ in $(\Sigma, f)$. This proves part~(ii).

For the proof of part~(iii), let us define  Boolean vector functions $f$ and $h$ on four-dimensional Boolean vectors
$s = (s_1, s_2, s_3, s_4)$ as follows:

$$f(s) = (s_1 \wedge s_2, s_1 \wedge s_3, s_2 \wedge s_4, s_3 \wedge s_4),$$
$$h(s) = (s_1 \vee s_2, s_1 \vee s_3, s_2 \vee s_4, s_3 \vee s_4).$$

Table~\ref{tab1} shows the values of $f, h, h \circ f$, and $f \circ h$.

\begin{center}
\begin{tabular}{| c | c | c | c | c |} \hline
$s$ & $f(s)$ & $h(s)$ & $h\circ f(s)$ & $f \circ h$
\\ \hline
1111 & 1111 & 1111 & 1111 & 1111\\
1110 & 1100 & 1111 & 1110 & 1111\\
1101 & 1010 & 1111 & 1101 & 1111\\
1100 & 1000 & 1110 & 1100 & 1100\\
1011 & 0101 & 1111 & 1011 & 1111\\
1010 & 0100 & 1101 & 1010 & 1010\\
1001 & 0000 & 1111 & 0000 & 1111\\
1000 & 0000 & 1100 & 0000 & 1000\\
0111 & 0011 & 1111 & 0111 & 1111\\
0110 & 0000 & 1111 & 0000 & 1111\\
0101 & 0010 & 1011 & 0101 & 0101\\
0100 & 0000 & 1010 & 0000 & 0100\\
0011 & 0001 & 0111 & 0011 & 0011\\
0010 & 0000 & 0101 & 0000 & 0010\\
0001 & 0000 & 0011 & 0000 & 0001\\
0000 & 0000 & 0000 & 0000 & 0000\\
 \hline
 \end{tabular}\label{tab1}
\end{center}

Let
$$F = \{1111, 1110, 1101, 1100, 1011, 1010, 0111, 0101, 0011, 0000\},$$
$$H = \{1111, 1100, 1010, 1000,  0101, 0100, 0011, 0010, 0001, 0000\}.$$
As Table~\ref{tab1} shows, $h \circ f$ is the identity on $F$ and
$f \circ h$ is the identity on $H$.

Let $L$ be a positive integer divisible by eight, and let $p:=L/4$.
Write $[L]$ as a disjoint union of blocks of four consecutive integers $i(1,r), i(2,r), i(3,r), i(4,r)$ for $r \in [p]$.  Call a Boolean vector $s \in \{0, 1\}^{[L]}$ \emph{$L$-compliant} if

\begin{itemize}
\item[(a)] $(s_{i(1,r)}, s_{i(2,r)}, s_{i(3,r)}, s_{i(4,r)}) \in F$ for $1 \leq r \leq p/2$,
\item[(b)] $(s_{i(1,r)}, s_{i(2,r)}, s_{i(3,r)}, s_{i(4,r)}) \in H$ for $p/2 < r \leq p$, and
\item[(c)] $s$ takes the value $1$ exactly $L/2$ times.
\end{itemize}

\begin{lemma}\label{compliantlem}
Let $c_1 < 10^{1/4}$.  Then there exist a positive integer $\ell$ and a positive integer $L$ that is a multiple of eight such that $2^\ell > c_1^L$ and the number of $L$-compliant Boolean vectors is larger than $2^\ell$.
\end{lemma}

\noindent
\textbf{Proof:} Let $L$ be a positive integer that is an integer multiple of 16, and let $V$ be the set of Boolean vectors  $s \in \{0,1\}^L$  that
satisfy conditions~(a) and~(b) above.  Since $|F| = |H| = 10$, it is clear that $|V| = 10^{L/4}$.

For each $s \in V$ define the \emph{signature of $s$} as $\sigma(s) = (\sigma_1(s), \dots , \sigma_6(s))$, where

\noindent
$\sigma_1(s) = |\{r: \ 1 \leq r \leq p/2 \ \& \ (s_{i(1,r)}, s_{i(2,r)}, s_{i(3,r)}, s_{i(4,r)}) = (1111)\}|$,

\noindent
$\sigma_2(s) = |\{r: \ 1 \leq r \leq p/2 \ \& \ (s_{i(1,r)}, s_{i(2,r)}, s_{i(3,r)}, s_{i(4,r)}) = (0000)\}|$,

\noindent
$\sigma_3(s) = |\{r: \ 1 \leq r \leq p/2 \ \& \ (s_{i(1,r)}, s_{i(2,r)}, s_{i(3,r)}, s_{i(4,r)}) \in \{(1110), (1101), (1011), (0111)\}\}|$,

\noindent
$\sigma_4(s) = |\{r: \ p/2 < r \leq p \ \& \ (s_{i(1,r)}, s_{i(2,r)}, s_{i(3,r)}, s_{i(4,r)}) = (1111)\}|$,

\noindent
$\sigma_5(s) = |\{r: \ p/2 < r \leq p \ \& \ (s_{i(1,r)}, s_{i(2,r)}, s_{i(3,r)}, s_{i(4,r)}) = (0000)\}|$,

\noindent
$\sigma_6(s) = |\{r: \ p/2 < r \leq p \ \& \ (s_{i(1,r)}, s_{i(2,r)}, s_{i(3,r)}, s_{i(4,r)}) \in \{(1000), (0100), (0010), (0001)\}\}|$.

Let $\sigma^{max} = (1/16, 1/16, 1/4, 1/16, 1/16, 1/4)$.  Well-known properties of binomial coefficients imply that the inequality

\begin{equation}\label{sigineq}
|\{s \in V: \, \sigma(s) = \sigma\}| \leq |\{s \in V: \, \sigma(s) = \sigma^{max}\}|
\end{equation}

holds for any possible signature $\sigma$.
Moreover, observe that if $s \in V$ and $\sigma(s) = \sigma^{max}$, then $s$ takes the value 1 exactly $L/2$ times, and hence $s$ is
$L$-compliant.  Since the total number of possible signatures is bounded from above by $(L/4+1)^6$, it follows from~(\ref{sigineq}) that the total number $M$ of $L$-compliant Boolean vectors satisfies the inequality

$$
M \geq \frac{10^{L/4}}{(L/4+1)^6}.
$$

Notice that $\lim_{L \rightarrow \infty} \ L\ln 10^{1/4} - 6 \ln (L/4+1) - L \ln c_1 = \infty$.

Thus for sufficiently large $L$ we can find a positive integer $\ell$ with
$$ L\ln 10^{1/4} - 6 \ln (L/4+1) > \ell \ln 2 > L \ln c_1,$$

and the lemma follows. $\Box$
\bigskip

Now fix $c_1 < 10^{1/4}$ and let $L, \ell$ be as in Lemma~\ref{compliantlem}.  Build an $n$-dimensional Boolean system
$(\Pi, g^-)$ as in the proof of Theorem~\ref{theorem counterexample}(i), but with the following modifications:

\begin{itemize}
\item The blocks $S_i$ will have length $L$ as before, but the set $A$ will consist only of $L$-compliant vectors.
\item Proper initial states will be required to have only $L$-compliant vectors on each $S_i$.
\item Instead of requiring $S_{i}(t+1) = S_{i+1}(t)$ for $i \in [N]$ and implementing this dynamics by monic functions,
for $i \in [N-1]$ we only require $S_i(t+2) = S_{i+2}(t)$ and implement this dynamics as follows: Let $S_i$ be partitioned into blocks
$b_{i,1}, \ldots , b_{i,{L/4}}$
of four Boolean values each, with $b_{i,r}(t) \in F$ for $r \leq L/8$ and $b_{i,r} \in H$ for $L/8 < r \leq L/4$.
Define $b_{i,r}(t+1) = h(b_{i+1, r + L/8}(t))$
 for $r \leq L/8$ and $b_{i,r}(t+1) = f(b_{i+1, r - L/8}(t))$
 for $L/8 < r \leq L/4$.
\end{itemize}

This construction is possible by Lemma~\ref{compliantlem} and the observations on the functions $f, h$ made above, and the exact same argument as in the proof of Theorem~\ref{theorem counterexample}(i) shows that each proper state of $(\Pi, g^-)$ belongs to an orbit of length $\geq c_2^n$,
where $c_2$ is a constant that depends only on $L$ and $\ell$
and satisfies $c_1 < c_2 < 10^{1/4}$.
It is also straightforward to verify that the system is bi-quadratic and cooperative.

However, the system may not yet be strictly quadratic.  We may still need to implement the dynamics $S_{N}(t+1) = S_{N+1}(t)$ by monic functions
and assume wlog that $\mu$ is even to assure that we have an exact copy of a previous value for $S_{N+1}$ when it is read as an input.
More importantly, some of the regulatory functions in $B \cup D$ will be monic (see Figure~\ref{figure switch} and Figure~\ref{figure cooperative}).  However, the number $U$ of nodes with indegree 1 is bounded by a number $U(L)$ that depends only on $L$, regardless of $n$.

\begin{lemma}\label{upperlem}
Suppose $(\Pi, g)$ is a bi-quadratic, $0.25$-biased, $n$-dimensional Boolean network with exactly $U$ monic regulatory functions and an orbit of length $c^n$.
Then there exists a strictly quadratic, bi-quadratic, $0.25$-biased,
$U + n$-dimensional
Boolean network $(\Sigma, f)$ with an orbit of length $c^n$.  Moreover, if $(\Pi, g)$ is cooperative, then we can also require that $(\Sigma, f)$ be cooperative.
\end{lemma}

\noindent
\textbf{Proof:} Let $(\Pi, g)$, $U$ be as in the assumption.
Since the sum of indegrees in a directed graph is equal to the sum of outdegrees, the number  $U_1$ of variables of $(\Pi, g)$ with outdegree 1 and the number $U_0$ of variables with outdegree 0 are such that $U = 2U_0 + U_1$.   Let $\{u_1, \ldots , u_U\}$ be the nodes
of $(\Pi, g)$
with indegree 1, let $\{o_1, \ldots , o_{U_0}\}$ be the nodes with outdegree $0$, and let $\{p_{2U_0 + 1}, \ldots , p_{2U_0+U_1}\}$ be the nodes with outdegree 1. We can construct $(\Sigma , f)$ from $(\Pi, g)$ by adding a set
$I = \{i_1, \ldots i_U\}$ of dummy nodes to the system as follows:

\begin{itemize}
\item $f_{u_r} = g_{u_r} \wedge s_{i_r}$ for $r \in [U]$, where $g_{u_r}$ denotes the (monic) $u_r$-th component of $g$,
\item $f_{i_r} = s_{i_r} \vee s_{o_r}$ for $r \in [U_0]$,
\item $f_{i_r} = s_{i_r} \vee s_{o_{r + U_0}}$ for $U_0 < r \leq 2U_0$,
\item $f_{i_r} = s_{i_r} \vee s_{p_r}$ for $2U_0 < r \leq U$.
\end{itemize}

Leaving the remaining regulatory functions unchanged, we obtain a  bi-quadratic, strictly quadratic system $(\Sigma, f)$ which is cooperative whenever $(\Pi, g)$ is.
If the system starts in an initial state $s(0) = (s_1(0), \ldots , s_n(0), s_{i_1}(0), \ldots , s_{i_U}(0))$ with $(s_{i_1}(0), \ldots ,
s_{i_U}(0)) = (1, \ldots 1)$, then we will have $(s_{i_1}(t), \ldots , s_{i_U}(t)) = (1, \ldots , 1)$ along the trajectory, and the dynamics on the
 original $n$ variables remains unchanged.  In particular, if
$(s_1(0), \ldots , s_n(0))$ belongs to an orbit of $(\Pi, g)$ of length $c^n$, then $s$ will belong to an orbit of $(\Sigma , f)$ of the same length. $\Box$

Now extend $(\Pi, g^-)$ to an $n+U$ dimensional system as in Lemma~\ref{upperlem}.  The dimension of the extended system is bounded by $n + U(L)$.  Thus if we choose $n$ sufficiently large relative to $U(L)$ so that
$c_2^n > c_1^{n+U(L)}$, we obtain the conclusion of
Theorem~\ref{theorem counterexample}(iii). $\Box$
\bigskip

Theorem~\ref{theorem counterexample}(iii) is the special case $\alpha = 1$ of the following more general result:

\begin{proposition}\label{1(iii)prop}
Let $\alpha, c$ be constants with $0\leq \alpha \leq 1$ and $1 < c < 10^{\alpha/4}2^{1-\alpha}$.  Then for all sufficiently large $n$ there exist $n$-dimensional bi-quadratic cooperative Boolean networks that are $c$-chaotic and are not $(\alpha n, n)$-Turing systems.
\end{proposition}

\noindent
\textbf{Proof:} Let $\alpha , c$ be as in the assumption, let $c_1, c_2, c_3$ be such that $1 < c_1 < 10^{1/4}$, $1 < c_2 < c_3 < 2$, and $c = c_1^{\alpha}c_2^{1-\alpha}$.  Theorem~\ref{theorem counterexample} already covers the case $\alpha = 1$, so assume $\alpha < 1$.  We need to construct $n$-dimensional systems with the required properties that have $> \beta n$ strictly quadratic nodes.  We can find $L_1, L_2, \ell_1, \ell_2$ such that $L_1, \ell_1$ satisfy the conclusion of Lemma~\ref{compliantlem},
$$\binom{L_2}{L_2/2} > 2^{\ell_2} > c_3^{L_2},$$ and
$$\gamma > \frac{L_1}{L_1 + L_2} > \alpha,$$
where
$$c_3^{1 - \gamma} > c_2^{1-\alpha} .$$

Now construct a cooperative, bi-quadratic, $n$-dimensional Boolean network as in the proof of Theorem~\ref{theorem counterexample} with blocks
$S_i$ for $i \in [N+1]$ of length $L_1 + L_2$ each in such a way that that the values of the first $L_1$ variables in $S_i(t+1)$ will be computed
from the variables in the first $L_1$ entries of $S_{i+1}(t)$ as in the proof of Theorem~\ref{theorem counterexample}(iii), and the remaining $L_2$ variables of $S_i(t+1)$ will simply be copied from the corresponding variables of $S_{i+1}(t)$.  The proof of Theorem~\ref{theorem counterexample} (with some very minor adjustments) shows that for sufficiently large $n$ relative to $L_1 + L_2$ the resulting system will be bi-quadratic, cooperative, will have more than $\alpha n$ strictly quadratic nodes, and will have an orbit of length $\geq 2^{(\ell_1 + \ell_2)N}$, where
$N \approx \frac{n}{L_1 + L_2}$.  By our choice of $L_1, L_2, \ell_1, \ell_2$ we will have

$$2^{\frac{\ell_1 + \ell_2}{L_1 + L_2}} > c_1^{\frac{L_1}{L_1 + L_2}}c_3^{\frac{L_1}{L_1 + L_2}} > c_1^\alpha c_3^{1 - \gamma} > c_1^{\alpha}c_2^{1-\alpha} =  c,$$
and the result follows. $\Box$

\section{Appendix B: Proof of Theorem~\ref{GenTthm}}

A proof of Theorem~\ref{GenTthm} was reported in~\cite{PartII}. Here we give a slightly improved version of this proof.

We will prove Theorem~\ref{GenTthm} in two stages.  In the first stage of the proof we will show that very large subsets of the state space $\Pi$ of an $n$-dimensional Boolean system must be \emph{balanced} in a sense that will be defined shortly.  In the second stage of the proof we will show that if $\cS$ is the set of states in an orbit of an $\eps$-biased $(b,r)$-Boolean system and $\cS$ is sufficiently balanced, then only a small fraction of the regulatory functions can be non-monic.

\subsection{Balanced subsets of the state space}

Let $\Pi = \{0,1\}^{[n]}$ be the state space of an $n$-dimensional Boolean system.    Let $\cS = \{s^\ell: \, \ell \in \cL\}$ be a sequence of (not necessarily pairwise distinct) elements of $\Pi$.   If the elements of $\cS$ happen to be pairwise distinct, then we will speak of $\cS$ being a
\emph{subset} of $\Pi$.

To illustrate the key idea of this section, let $i \in [n]$ and consider the ratio

$$\zeta_i (\cS) = \frac{|\{\ell \in \cL: \, s^\ell_i = 1\}|}{|\cL|}.$$

If $\beta, \gamma > 0$, then we will say that $\cS$ is \emph{$\beta$-$\gamma$-$1$-balanced} if $|\{i \in [n]: \ |\zeta_i(\cS) - 0.5| \geq \gamma\}| < \beta n$.

More generally, let $r \in [n]$ and $\sigma : [r] \rightarrow \{0,1\}$.  For $r$-element subsets $I = \{i_1, \ldots , i_r\}$ of $[n]$ with
$i_1 < \dots < i_r$
we define ratios
$\xi_{I}^{\sigma}(\cS)$ as follows:

$$\xi^{\sigma}_{I} (\cS) = \frac{|\{\ell \in \cL: \, \forall u \in [r] \ s_{i_u}^\ell = \sigma(u)\}|}{|\cL|}.$$

Define

$$\zeta^*_I(\cS) = \max \{2^{-|I|} - \xi^{\sigma}_{I}(\cS): \sigma \in  \{0,1\}^{[r]} \}.$$

If $\beta, \gamma > 0$, then we will say that $\cS$ is \emph{$\beta$-$\gamma$-$r$-balanced} if for every family $P$ of pairwise disjoint
subsets $I$ of $[n]$ with $|\bigcup P| \geq \beta n$ and $1 \leq |I| \leq r$ for each $I \in P$ there exists $I \in P$ such that
$\zeta^*_I(\cS) < \gamma$.

We will prove the following.

\begin{lemma}\label{balancedlemma}
Let $r$ be a positive integer, $\beta, \gamma > 0$ and assume $\gamma < 2^{-r}$. Let

$$\lambda(\gamma, r) = \left(\frac{1- 2^{-r}}{1- 2^{-r}+\gamma}\right)^{1- 2^{-r}+\gamma}\left(\frac{2^{-r}}{2^{-r}-\gamma}\right)^{2^{-r}-\gamma},$$
and let $c$ be a constant such that
$$c > 2(\lambda(\gamma, r))^\beta.$$
Then  for sufficiently large $n$, every subset $\cS$ of $\{0,1\}^{[n]}$ of size $\geq c^n$ is  $\beta$-$\gamma$-$r$-balanced.
\end{lemma}

\noindent
\textbf{Proof:}  Let $\beta, \gamma, r$ be as in the assumptions, and assume throughout this argument that~$n$ is a sufficiently large positive
integer.  Let $\varrho > 0$, let $1 < c < 2$, and let $\delta$ be such that
$1 + \varrho\gamma < \delta < 1 + 2\varrho\gamma$ and $\delta c^n$ is an integer.
Let us assume that $\cS = \{s^\ell: \, \ell \in [\delta c^n]\}$ is a sequence of randomly and independently (with replacement) chosen states in
 $\{0,1\}^{[n]}$ of length
$\delta c^n$.
We will treat  $\xi_I^\sigma$ and $\zeta^*_I$ as random variables and temporarily suppress their dependence on $\cS$ in our notation.

Let $v \in [r]$. For fixed $I = \{i_1, \ldots , i_v\}$ with $i_1 < \dots < i_v$ and $\sigma \in \{0, 1\}^{[v]}$ we define

$$\eta_I^\sigma = \frac{\sum_{\ell=1}^{\delta c^n} \eta_{I\ell}^\sigma}{\delta c^n},$$

where $\eta_{I\ell}^\sigma = 0$ if $ s^\ell(i_u) = \sigma (u)$ for all $u \in [v]$, and $\eta_{I\ell}^\sigma = 1$ otherwise.

Clearly, the mean value of $\eta^\sigma_{I}$
is $E(\eta^\sigma_{I}) = 1 - 2^{-v}$.  Note that $2^{|I|} - \xi^\sigma_I \leq \eps$ iff $\eta^\sigma_{I} -E(\eta^\sigma_{I}) \geq \eps$, and hence
$\zeta^*_I \geq \eps$ iff $\eta^\sigma_{I} -E(\eta^\sigma_{I}) \geq \eps$ for at least one
$\sigma \in \{0,1\}^{[v]}$.

We want to estimate $Pr(\eta^\sigma_{I} -E(\eta^\sigma_{I}) \geq \eps)$ for any given fixed $\eps > 0$.
Note that the random variables $\eta_{I\ell}^{\sigma}$ take values in the interval $[0,1]$ and are independent.  This allows us to use the
following inequality of \cite{Hoeffding} (see also \cite{Okamoto, Chernoff} for the special case we are considering here).

\begin{lemma}\label{Jansoneq}
Let $X_1, X_2, \dots, X_m$ be independent random variables such that $0 \leq X_i \leq 1$ for $i \in [m]$ and let
$X = (X_1 + \dots + X_m)/m$.  Let $\mu = E(X)$ and let $0< \eps < 1-\mu$.   Then

\begin{equation}\label{Hoeffeqn}
Pr(X - \mu \geq \eps) \leq \left(\left(\frac{\mu}{\mu+\eps}\right)^{\mu+\eps}\left(\frac{1-\mu}{1 - \mu-\eps}\right)^{1 - \mu-\eps}\right)^m
\leq e^{-2\eps^2m}.
\end{equation}
\end{lemma}

We will assume until further notice that $\eps < 2^{-v}$ and thus satisfies the assumptions of~(\ref{Hoeffeqn}).
Both bounds in~(\ref{Hoeffeqn}) are of the form $\lambda^m$ for some $0 < \lambda \leq e^{-2\eps^2} < 1$.  For the moment, assume that
$\lambda$ is such such a constant, and let
$m = \delta c^n$.  Now it follows from~(\ref{Hoeffeqn}) that

$$Pr(\eta_{I}^\sigma - 1 + 2^{-v} \geq \eps) \leq \lambda^{\delta c^n}.$$

This implies the following estimate for $\zeta^*_I$:

$$Pr(\zeta^*_I  \geq \eps) \leq 2^v \lambda^{\delta c^n}.$$

Now fix $k < n$ and consider $k$ pairwise disjoint subsets $I_1, \ldots , I_k$ of cardinality $\leq r$ each.
The random variables $\zeta^*_{I_1}, \ldots , \zeta^*_{I_k}$ are independent.  It follows that

$$Pr(\forall m \in [k] \ \zeta^*_{I_m} \geq \eps) \leq 2^{rk} \lambda^{k\delta c^n}.$$

Let $k = \beta n$ and let $A$ be the event that there exists a family $P$ of pairwise disjoint
subsets~$I$ of $[n]$ with $|\bigcup P| \geq \beta n$ and $1 \leq |I| \leq r$ for each $I \in P$ such that $\zeta^*_{I} \geq \eps$ for each
 $I \in P$.  The number of eligible families $P$ is bounded from above by $\binom{n}{r}^{\beta n} < n^{r\beta n}$.
 Thus the probability of the event $A$ can be estimated as

$$Pr(A) < (2n)^{r\beta n}\lambda^{\beta n \delta c^n}.$$

Now note that by Stirling's formula the number of subsets of $\Pi$ of size $c^n$ satisfies

$$\binom{2^n}{c^n} < \frac{2^{nc^n}}{c^n!} < \frac{1}{2}\frac{2^{nc^n} e^{c^n}}{c^{nc^n}} = \frac{1}{2}\left(\frac{2e^{\frac{1}{n}}}{c}\right)^{nc^n}.$$

Moreover, note that

$$\lim_{n \rightarrow \infty} (2n)^{\frac{r\beta}{c^n}} = 1.$$

Thus for

\begin{equation}\label{cestimate}
c > 2\lambda^{\beta\delta}
\end{equation}

and $n$ sufficiently
large, we will have

$$(2n)^{\frac{r\beta}{ c^n}}\lambda^{\beta\delta} < \left(\frac{2e^{\frac{1}{n}}}{c}\right)^{-1}.$$

This in turn implies that for sufficiently large $n$ and $c$ as in~(\ref{cestimate})

\begin{equation}\label{punch1}
Pr(A) < (2n)^{r\beta n}\lambda^{\beta n\delta c^n} =  \left((2n)^{\frac{r\beta}{c^n}}\lambda^{\beta \delta}\right)^{n c^n}
< (\frac{2e^{\frac{1}{n}}}{c})^{-n  c^n} < \frac{1}{2\binom{2^n}{c^n}}.
\end{equation}

Now let us fix $\eps$ such that $0 < \eps < \gamma$.  Since $\gamma < 2^{-r}$, the assumptions of Lemma~\ref{Jansoneq}
will be satisfied for this choice of $\eps$. Let $B = B(\cS)$ be the set of the first $c^n$ pairwise distinct elements of the sequence $\cS$, if in fact
$\cS$ has at least $c^n$ pairwise distinct elements, and let~$B$ be undefined otherwise.  Let us make a few observations:

\begin{enumerate}
\item Let $N = \{\ell \in [\delta c^n]: \ \exists 1 \leq j < \ell \ s^j = s^\ell\}$ be the number of entries in $\cS$ that duplicate
a previous entry.  Note that $B$ is defined iff $N \leq (\delta - 1) c^n$.  In particular, by the choice of $\delta$, the set
$B$ is defined as long as  $N \leq \varrho\gamma c^n$.

\item Note that the expected value of $N$ can be estimated, for sufficiently large $n$, fixed $c < 2$, and $0 < \varrho < \frac{2-c}{2c\gamma}$, as

$$E(N) \leq \sum_{\ell \in [\delta c^n]} \frac{\ell-1}{2^{n}} < \frac{\delta^2 c^{2n}}{2^n} = o(1) c^n.$$

In particular, $E(N) < \frac{\delta - 1}{2}c^n$.

\item Now it follows from Markov's Inequality

$$\frac{\delta - 1}{2}c^n > E(N) \geq Pr(N > (\delta -1)c^n)(\delta-1)c^n$$
 that for fixed $c$ and sufficiently large $n$, the set $B$ will be defined with probability
$> 0.5$.

\item Assume $B$ is defined.  Observe that for  each subset $I$ of $[n]$ and
$\sigma \in \{0, 1\}^{[|I|]}$ we have

\begin{equation}\label{BSeqn}
\frac{\eta^\sigma_I(B)}{\delta} \leq \eta^\sigma_I(\cS)\leq \frac{\eta^\sigma_I(B) + \delta - 1}{\delta}.
\end{equation}

The first inequality in~(\ref{BSeqn}) turns into equality if $\eta_{I\ell}^\sigma = 0$ whenever $s^\ell$ is outside of~$B$;
the second inequality in~(\ref{BSeqn}) turns into equality if $\eta_{I\ell}^\sigma = 1$ whenever $s^\ell$ is outside of~$B$.
It follows from the relationship between the $\eta^\sigma_I(\cS)$'s and $\zeta^*_I(\cS)$ that
$$ \frac{\zeta^*_I(B)}{\delta} \leq \zeta^*_I(\cS)\leq \frac{\zeta^*_I(B) + \delta - 1}{\delta}.$$
By choosing $\varrho$ sufficiently close to $0$, we can choose $\delta$ as close to 1 as we need, and
our choice of $\eps < \gamma$  implies that for $\delta$ sufficiently close to 1 the inequality $\zeta^*_I(B) \geq \gamma$  will
 imply the inequality $\zeta^*_I(\cS) \geq \eps$.
\end{enumerate}

But if there is \emph{any} subset $B$ of size $c^n$ of $\Pi$ that is not $\beta$-$\gamma$-$r$ balanced, then
 this subset will be exactly as likely to be equal to $B(\cS)$ as any
other subset of $\Pi$ of the same size. By point 3 above, the probability that $B(\cS)$ exists is greater than $0.5$, and thus the probability that
$B(\cS)$ exists and is equal to $B$ must be
at least $0.5 \binom{2^n}{c^n}^{-1}$.
But point~4 above implies that if $B$ is not $\beta$-$\gamma$-$r$ balanced, then $B(C) = B$ implies that the event $A$
has occurred, with contradicts inequality~(\ref{punch1}).

We derived the contradiction under the assumption that $c$ satisfies inequality~(\ref{cestimate}).  Now assume
$$c > 2(\lambda(\gamma, r))^\beta$$ as in the assumption of the lemma.  Then we can choose
$\eps$ sufficiently close to $\gamma$ and $\lambda = \lambda(\eps, r)$
so that inequality~(\ref{cestimate}) will hold as well for any $\delta > 1$. By choosing
$\delta$ sufficiently close to one we will get a contradiction whenever $B$ exists and satisfies $\zeta^*_I(B) \geq  \gamma$.
This proves Lemma~\ref{balancedlemma}. $\Box$

\subsection{Systems with balanced orbits}

\begin{lemma}\label{orbitlem}
Let $b,r$ be positive integers, let $0 <  \eps, \tau < 0.5$, let $(\Pi, g)$ be an $n$-dimensional Boolean system, and let $\cS$ be an orbit of
$(\Pi, g)$.
Let $k \in [n]$ be such that the bias~$\Lambda$ of $g_k$ satisfies $|\Lambda - 0.5| \geq \eps$, and let $I$ be the set of input variables
of $g_k$.  Then either $\zeta^*_I(\cS) \geq \frac{\tau}{2^{|I|}}$ or $\zeta^*_{\{k\}}(\cS) \geq (1-\tau)\eps - \frac{\tau}{2}$.
\end{lemma}

\noindent
\textbf{Proof:} Assume wlog that $\Lambda \geq 0.5 + \eps$; the proof in the case when $\Lambda \leq 0.5 - \eps$ is symmetric.
Suppose that $\zeta^*_I(\cS) < \frac{\tau}{2^{|I|}}$.  Then there exists a subset $\cS^* \subseteq \cS$ with $|\cS^*| \geq (1 - \tau)|\cS|$ such that $\eta_I^\sigma(\cS^*) = 2^{-|I|}$ for each $\sigma \in \{0,1\}^I$.
We conclude that
\begin{equation*}
\begin{split}
&|\cS|\zeta_k = |\{s \in \cS: \ s_k = 1\}| = |\{s \in \cS: \ g_k(s) = 1\}| \geq |\{s \in \cS^*: \ g_k(s) = 1\}|\\
&\geq (1- \tau)|\cS|\Lambda \geq \left(1- \tau\right)|\cS|\left(0.5 + \eps\right) > |\cS|\left(0.5 + (1-\tau)\eps - \frac{\tau}{2}\right),
\end{split}
\end{equation*}
and the inequality  $\zeta^*_{\{k\}} \geq (1-\tau)\eps - \frac{\tau}{2}$ follows. $\Box$

\begin{lemma}\label{cyclelemma}
Let  $(\Pi, g)$ be an $n$-dimensional $\eps$-biased $(n,r)$-Boolean system, let $0 < \tau < \frac{\eps}{\eps + 0.5}$, let $\gamma = \frac{\tau}{2^{r}}$, $\gamma^* = (1-\tau)\eps - \frac{\tau}{2}$, and let $\beta, \beta^* > 0$.  Assume $\cS$ is the set of states in an orbit of $(\Pi, g)$ so that $\cS$ is both $\beta$-$\gamma$-$r$-balanced and
$\beta^*$-$\gamma^*$-$1$-balanced.  Then  there exists a subset $J \subseteq [n]$ of size $|J| < (\beta  + r \beta^*) n$ with the property that every non-monic regulatory function $g_k$ has at least one input variable in $J$.
\end{lemma}

\noindent
\textbf{Proof:} Let $K = \{ k \in [n]: \ \zeta^*_{\{k\}}(\cS) \geq \gamma^*\}$. The assumption on $\cS$ implies that $|K| < \beta^* n$.

Let $J_0$ be the set of inputs of the variables in $K$.  Then $|J_0| < r\beta^* n$.

Let $K^+ = [n] \backslash K$ and let $\{k_1, \ldots , k_p\} \subseteq K^+$ be a set of variables maximal with respect to the property that
$g_{k_q}$ is non-monic for every $q \in [p]$ and the sets $I_q$ of inputs of $g_{k_q}$ are pairwise disjoint.  Let $J_1 = \bigcup_{q \in [p]} I_q$.

By Lemma~\ref{orbitlem} and the choice of $K^+$, for each $q \in [p]$ we must have $\zeta^*_{I_q}(\cS) \geq \frac{\tau}{2^{r}}$.  Thus the assumption on $\cS$ implies that $|J_1| < \beta n$.

On the other hand, by maximality of $\{k_1, \ldots , k_p\}$, every non-monic regulatory function~$g_k$ must have at least one input in the set
$J:= J_0 \cup J_1$, and the lemma follows. $\Box$
\bigskip

Now let $(\Pi,g), \eps, \alpha, b, r$ be as in the assumptions of Theorem~\ref{GenTthm}, let $\gamma, \gamma^*$ be as in the assumptions of Lemma~\ref{cyclelemma}, and assume that $\beta, \beta^* > 0$ satisfy

\begin{equation}\label{beteqn}
\beta + r\beta^* = \frac{\alpha}{b}.
\end{equation}

Let $\lambda(\gamma, r), \lambda(\gamma^*, 1)$ be as in Lemma~\ref{balancedlemma}.  Then we will have

\begin{equation}\label{ceps}
c(\eps, \alpha, b, r) \leq \max\{2(\lambda(\gamma, r))^{\beta}, 2(\lambda(\gamma^*, 1))^{\beta^*}\}.
\end{equation}

To see this, assume $n$ is sufficiently large and $\cS$ is an orbit of $(\Pi,g)$ of length at least $c^n$, where
$c$ exceeds the right-hand side of~(\ref{ceps}).  Then Lemma~\ref{balancedlemma} implies that
$\cS$ is $\beta$-$\gamma$-$r$-balanced and $\beta^*$-$\gamma^*$-$1$-balanced
and thus satisfies the assumptions of Lemma~\ref{cyclelemma}.  Let~$J$ be as in the conclusion of
Lemma~\ref{cyclelemma}.  Note that at most $b|J| < \alpha n$ regulatory functions can have inputs in $J$, and it follows that $(\Pi, g)$ is an
$(\alpha n, n)$-Turing system.

Note that by the second inequality in~(\ref{Hoeffeqn}) we will in particular have

\begin{equation}\label{cepsw}
c(\eps, \alpha, b, r) \leq \max\{2e^{-2\gamma^2\beta}, 2e^{-2(\gamma^*)^2\beta^*}\}.
\end{equation}

This concludes the proof of Theorem~\ref{GenTthm}. $\Box$

\section{Appendix C: Numerical Estimates for $c(\eps, \alpha, b, r)$}

We formulated  Theorem~\ref{GenTthm} as a qualitative result about existence of a constant.
In this section we will use the notation $c(\eps, \alpha, b,r)$ as shorthand for the
smallest
real number for which the conclusion of Theorem~\ref{GenTthm} holds and $C(\eps, \alpha, b,r)$ for the upper bound for $c(\eps, \alpha, b,r)$ that we get from our proof of the theorem.

To arrive at more precise estimates of
$c(\eps, \alpha, b, r)$, we defined $\gamma = \frac{\tau}{2^r}$, $\gamma^* = (1-\tau)\eps - \frac{\tau}{2}$, where
$\tau$ is as in the assumptions of Lemma~\ref{cyclelemma}, and
wrote a simple MatLab program for numerically exploring the values of the right-hand side of~(\ref{ceps}) for
$\tau \in (0, \frac{\eps}{\eps+0.5})$ and $\beta^* \in (0, \frac{\alpha}{br})$. Note that the value of $\beta$ is not a free parameter as it is
given by~(\ref{beteqn}).

For the cases that we numerically explored, we found almost perfect linear dependence of $C(\eps, \alpha, b,r)$ on $\alpha$.  In particular,

\begin{equation*}\label{cepslin}
\begin{split}
C(0.25, \alpha, 2, 2)  &\approx 2 - 0.0041\alpha,\\
C(0.25, \alpha, 3, 2)  &\approx 2 - 0.0027\alpha,\\
C(0.375, \alpha, 2, 3)  &\approx 2 - 0.0040\alpha,\\
C(0.25, \alpha, 2, 3)  &\approx 2 - 0.0021\alpha,\\
C(0.125, \alpha, 2, 3)  &\approx 2 - 0.0007\alpha,\\
C(0.375, \alpha, 3, 3)  &\approx 2 - 0.0027\alpha,\\
C(0.25, \alpha, 3, 3)  &\approx 2 - 0.0014\alpha,\\
C(0.125, \alpha, 3, 3)  &\approx 2 - 0.0004\alpha.
\end{split}
\end{equation*}

\begin{figure}[ht]
\centerline{\includegraphics[width=4in]{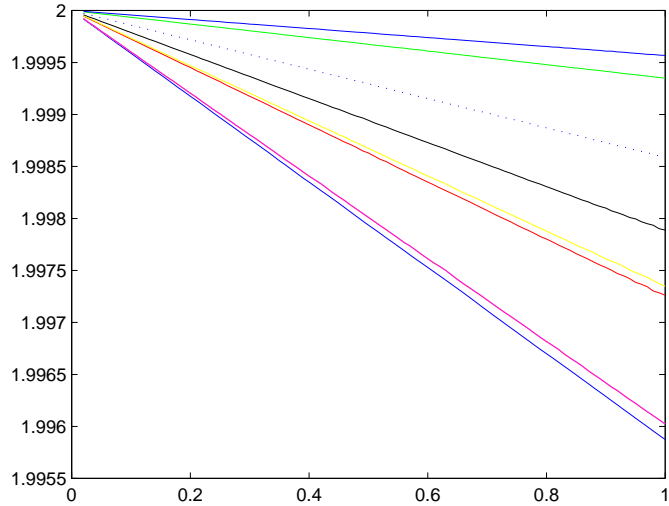} }
\caption{Dependence of $C(\eps, \alpha, b, r)$ on $\alpha$.
Solid blue: $\eps = 0.25, b=r=2$, Red: $\eps = 0.25, b=3, r=2$, Magenta: $\eps = 0.375, b = 2, r = 3$,
Black: $\eps = 0.25, b=2, r = 3$, Green: $\eps = 0.125, b=2, r = 3$, Yellow: $\eps = 0.375, b=r=3$,
Dotted Blue: $\eps = 0.25, b=r=3$, Dashed Blue: $\eps = 0.125, b=r=3$.}
\label{figure cees}
\end{figure}

Figure~\ref{figure cees} summarizes these findings.

The upper bounds $C(\eps, \alpha, b, r)$ appear to substantially overestimate the values $c(\eps, \alpha, b, r)$.  For example, notice that
$C(0.25, 1, 2, 2) \approx 1.9959$.  Note that for $c > c(0.25, 1, 2, 2)$ no strictly bi-quadratic $0.25$-biased Boolean network can be $c$-chaotic. Thus Theorem~\ref{theorem counterexample}(iii) gives a lower bound
$c(0.25, 1, 2, 2) \geq 10^{1/4} \approx 1.7783$.  We prove below that $10^{1/4}$ is in fact the correct value of $c(0.25, 1, 2, 2)$.

Let us call a vector $(g_1, \ldots , g_n)$ of Boolean functions on $\{0,1\}^{I}$ with $|I| = n$ a \emph{Boolean $n$-block} if each of the functions $g_i$ is strictly quadratic, has bias $\Lambda = 0.25$ or $\Lambda = 0.75$, and each $i \in I$ acts as input to exactly two among the functions
$g_1, \ldots , g_n$.  A Boolean $n$-block is \emph{minimal} if there is no proper subset $J \subset [n]$ such that the $g_j$'s for $j \in J$ form a Boolean $|J|$-block on some proper subset of $I$.

For example, let us consider a Boolean $n$-block $G = (g_1, \ldots , g_n)$ on $I = [n]$ with $n > 2$.  Wlog, $g_1$ has inputs $s_1, s_2$, and $s_1$ is also an input of $g_2$.  If $s_2$ were the other input of $g_2$, then $(g_1, g_2)$ would form a Boolean $2$-block on $[2]$, and $G$ would not be minimal since we assumed $n > 2$.  Thus we may wlog assume that the other input of $g_2$ is $s_3$.  Inductively arguing like this we can convince ourselves that if $G$ is minimal, then after a suitable renumbering of the input variables we can assume wlog that $g_i$ takes inputs
$s_i, s_{i+1}$ for all $i \in [n-1]$ and $g_n$ takes inputs $s_1, s_n$.

Let us define $R(n)$ for $n \geq 2$ as the maximal size of the range of a minimal Boolean $n$-block, and let $\varrho(n) = (R(n))^{1/n}$.

\begin{lemma}\label{rholem}
$\varrho(n) \leq 10^{1/4}$ for all integers $n \geq 2$.
\end{lemma}

\noindent
\textbf{Proof:} Let $(g_1, \ldots , g_n)$ be a minimal Boolean $n$-block, and assume wlog that
$g_i$ takes input variables $s_i, s_{i+1}$ for $i < n$ and $g_n$ takes input variables $s_n, s_1$.  Let the \emph{opening} of the block
be the vector of Boolean functions $(g_1, \ldots , g_n)$ where $g_1, \ldots , g_{n-1}$ are as before, but $g_n$ is now treated as a Boolean function with inputs $s_n, s_{n+1}$. We will not make a notational distinction between a Boolean $n$-block and its opening.   Note that the definition of a Boolean $n$-block implies that each $g_i$ is canalyzing in both variables.

Assume wlog that the canalyzed values of all $g_i$ are 1; if not, we can replace $g_i$ by $\neg g_i$  without altering the size of the range of $(g_1, \ldots , g_{n})$ on any given set of input vectors.

Let $(g_1, \ldots , g_n)$ be the opening of a minimal Boolean $n$-block.  Define

\begin{itemize}
\item $I_{00}$ as the set of input vectors $(s_1, \ldots s_{n+1})$ such that $s_1$ takes the value that does not canalyze $g_1$ and $s_{n+1}$ takes the value that does not canalyzes $g_{n}$,
\item $I_{01}$ as the set of input vectors $(s_1, \ldots s_{n+1})$ such that $s_1$ takes the value that does not canalyze $g_1$ and $s_{n+1}$ takes the value that canalyzes $g_{n}$,
\item $I_{10}$ as the set of input vectors $(s_1, \ldots s_{n+1})$ such that $s_1$ takes the value that  canalyzes $g_1$, and $s_{n+1}$ takes the value that does not canalyze $g_{n}$,
\item $I_{11}$ as the set of input vectors $(s_1, \ldots s_{n+1})$ such that $s_1$ takes the value that canalyzes $g_1$, and $s_{n+1}$ takes the value that canalyzes $g_{n}$.
\end{itemize}

Let $R_{ij}$ be the range of $(g_1, \ldots , g_{n})$ restricted to $I_{ij}$.

\begin{proposition}\label{cancanprop}
Let $n \geq 2$.  Then $|R_{01} \cup R_{10}| \leq |R_{00} \cup R_{11}|$.
\end{proposition}

\noindent
\textbf{Proof:} Let $r = (r_2, \ldots , r_{n-1})$ be a Boolean vector.
It suffices to show for $k \in [4]$ that if exactly $k$ among the vectors $irj$ for $i, j \in \{0,1\}$ belong to  $R_{01} \cup R_{10}$, then
at least $k$ among these vectors must belong to $R_{00} \cup R_{11}$.

For $k = 1$ this is obvious, because the vector $1r1$ always belongs to $R_{00} \cup R_{11}$.

For $k = 4$ it is vacuously true, since the output vector $0r0$ can never belong to $R_{01} \cup R_{10}$.

For $k = 2$, assume wlog that $0r1 \in R_{01}$, and let $s = (s_1, \ldots , s_{n+1})$ be a corresponding input vector.  If $s$ is obtained by flipping the value
of $s_{n+1}$ to $1-s_{n+1}$ and leaving $s$ otherwise unchanged, we get an input vector $s' \in I_{00}$ with output $0rj$ for some $j \in \{0, 1\}$, and it follows that both $1r1, 0rj \in R_{00} \cup R_{11}$.

For $k = 3$, let $s^1, s^2, s^3 \in I_{01} \cup I_{10}$ with outputs $0r1, 1r1, 1r0$ respectively.  Then $s^1 \in I_{01}, s^3 \in I_{10}$, and wlog $s^2 \in I_{10}$.  Again, flipping the last value of $s^1$ gives an input in $I_{00}$ with output $0rj$, and flipping the first value of
of $s^3$ gives an input in $I_{00}$ with output $ir0$.  If $i \neq 0$ or $j \neq 0$ we are done.  If not, consider the vectors
$(s^1_2, \ldots , s^1_{n})$ and $(s^3_2, \ldots , s^3_{n})$.  If there is some $i$ with $2 \leq i \leq n$ so that $s^1_i = s^3_i$, then we can form an input vector $(s^1_1, \ldots , s^1_i, s^3_{i+1}, \ldots , s^3_{n+1}) \in I_{00}$ with value $0r1$, and we are done.  If not, then we must have $(r_2, \ldots , r_{n-1}) = (1, \ldots, 1)$, since the non-canalyzed value can only be taken if both inputs are at their non-canalyzing values.  In this special case  we can consider two input vectors in $I_{00}$ where $s_i$ is the canalyzing value for $g_i$ for $i = 2, \ldots , n-1$ and $s_{n}$ is arbitrary.  The output vectors will be $jr0$ and $jr1$ for some $j \in \{0, 1\}$.  Since we need to consider this last case only if we have already found that $0r0 \in I_{00}$, we are done.  $\Box$
\bigskip

Now let us consider a minimal Boolean $n$-block $(g_1, \ldots , g_{n})$. but let us for the moment assume that $g_{n-1}$ takes inputs $s_{n-1}, s_n$.  The range of this block is equal to the range of its opening  restricted to inputs such that
$s_1 = s_{n+1}$.  If the canalyzing value of $s_1$ for $g_1$ is equal to the canalyzing value of $s_{n}$ for $g_{n}$, then the size of the range of the block
$(g_1, \ldots , g_{n})$ is equal to $|R_{00} \cup R_{11}|$; otherwise it is equal to $|R_{01} \cup R_{10}|$.  This gives the following:

\begin{corollary}\label{cancor}
Let $(g_1, \ldots , g_{n})$ be a minimal Boolean $n$-block
whose range is of maximum possible size.  Then we may wlog assume that each input variable has the same canalyzing value for both its output functions.
\end{corollary}

\noindent
\textbf{Proof:}  Suppose $(g_1, \ldots , g_{n})$ is a minimal Boolean $n$-block whose range is of maximum possible size, with minimum number $m$ of input variables with different canalyzing values for their two output functions.  Assume towards a contradiction that $m > 0$.  Without loss of generality, the first input variable has two different canalyzing values.  Consider $R_{ij}$ defined for its opening.   Then the range of the Boolean $n-1$-block $(g_1, \ldots , g_{n-1})$ has size $|R_{01} \cup R_{10}|$.  Replace $g_{n}$ by $f_{n}$ so that
$g_{n}(s_{1},s_{n}) = f_{n}(s_{1}, 1-s_{n})$ for all $s_{1}, s_n$. Note that this does not alter the definitions of the sets $R_{ij}$.  Thus we obtain a minimal $n$-block $(g_1, \ldots , g_{n-1}, f_n)$ whose range has size $|R_{00} \cup R_{11}|$, which still must be maximum by
Proposition~\ref{cancanprop}.   However, the
canalyzing value of $s_1$ for $g_1$ is now equal to the canalyzing value of $s_1$ for $f_{n}$, which contradicts the choice of $m$. $\Box$
\bigskip

Note that the size of the range of such a Boolean $n$-block $(g_1, \ldots , g_n)$ does not change if we replace one of the functions $g_i$ by $\neg g_i$ or if we simultaneously replace  functions $g_i, g_{i+1}$ with
$g_i\circ \neg s_i$ and $g_{i+1} \circ \neg s_i$.  Thus Corollary~\ref{cancor} implies that the maximum value for $R(n)$ is always attained by the minimal Boolean $n$-block $(g_1, \ldots , g_n)$ where $g_i = s_i \vee s_{i+1}$ for $i \in [n-1]$ and $g_n = s_1 \vee s_n$.
We will from now on assume that a $(g_1, \ldots ,g_n)$ is this particular $n$-block.  Note that, in particular, the canalyzing and canalyzed values will always be 1 in this case, for all $g_i$ and all input variables.

Note that an output vector $(r_1, \ldots , r_n)$ of $(g_1, \ldots , g_n)$ cannot contain an isolated 1, that is, there cannot be $i \in \{2, \ldots , n-1\}$ with
$s_i = 1$ and both $s_{i - 1} = 0 = s_{i + 1}$, or $s_1 = 1$ and $s_2 = s_ n = 0$ or $s_n = 1$ and $s_1 = s_{n-1} = 0$.
Let $a_n$ be the number of Boolean vectors of length $n$ without isolated ones.  The recursion $a_n=2a_{n-1}-a_{n-2}+a_{n-3}$ has been reported in
\cite{fib}, and it follows that $\lim_{n \rightarrow \infty} a_n^{1/n} = \lambda$, where  $\lambda = 1.7549$ is the real root of $x^3 -2x^2 +x -1$. For $n > 3$ the number $a_n$ has been reported online as $\lambda^{n+1}$ rounded to the nearest integer (see sequence A109377 in \cite{online}).  This already implies Lemma~\ref{rholem}.
At the time of this writing, no complete proof of the latter is
given in \cite{online}.  We include here an independent calculation of
$|R(n)|$ that uses a different recursion.

Fix $n \geq 2$ and consider the opening $(g_1, \ldots , g_n)$ of the minimal $n$-block of $\vee$-functions.  For $j \in \{0,1\}$ and $i \in [n]$ define:

\begin{itemize}
\item $p_j(i)$ as the size of the range of $(g_1, \ldots , g_i)$ restricted to inputs with $s_1 = j$ and $s_{i+1} = 1$,
\item $q_j(i)$ as the size of the range of $(g_1, \ldots , g_i)$ restricted to inputs with $s_1 = j$ and $s_{i+1} = 0$, and
\item $r_j(i)$ as the size of the range of $(g_1, \ldots , g_i)$ restricted to inputs with $s_1 = j$.
\end{itemize}

It is easy to see that $p_0(1) = p_1(1) = q_0(1) = q_1(1)$, $r_{1}(1) = 1$, and $r_{0} (1) = 2$.

Now assume $i \in [n-1]$ and $(r_1, \ldots , r_{i+1})$ is an output vector of $(g_1, \ldots , g_{i+1})$.  If $r_{i+1} \neq 1$,  then we must have $s_{i+1} = 0 = s_{i+2}$.  Similarly,
If $r_{i+1} = 1$, then we must have $s_{i+1} = 1$ or $s_{i+2} = 1$.  This observation leads to the
 the following recursive relationships:
\begin{itemize}
\item
$p_j(i+1) = r_j(i)$,

\item
$q_j(i+1) = p_j(i) + q_j(i)$,

\item
$r_j(i+1) = q_j(i) + r_j(i)$.
\end{itemize}

In other words, we have $(p_j(i+1), q_j(i+1), r_j(i+1))^T = M * (p_j(i), q_j(i), r_j(i))^T$, where

\begin{equation*}\label{M}
M = \left[
\begin{matrix}
0 & 0 & 1\\
1 & 1 & 0\\
0 & 1 & 1
\end{matrix}
\right]
\end{equation*}

The characteristic polynomial of $M$ is $\lambda^3 - 2\lambda^2 + \lambda - 1$, and
the eigenvalues are $\lambda_1 = 1.7549$,  $\lambda_2 = 0.1226 + 0.7449i$, $\lambda_3 =  0.1226 - 0.7449i$.
The normal eigenvector corresponding to $\lambda_1$ is $w_1 =(0.4140,  0.5484,   0.7265)^T$, and the vectors $w_2 = (-1.3117,
    0.8688, -0.1608)^T$, $w_3 = (0, -0.7376, 0.9771)^T$ span the eigenspace of eigenvalues $\lambda_2, \lambda_3$.  Note that the norm of $M$ on the subspace spanned by $w_2, w_3$ is $|\lambda_2| = 0.7549$.  The vector $(1,1,1)^T$ can be written in base $W = (w_1, w_2, w_3)$ as $u_1 = (1.7443, -0.2118, -0.3084)^T$, and the vector $(1,1,2)^T$ can be written in this base as $u_0 = (2.3107, -0.0331,  0.3234)^T$.
    Let $L = W^{-1}MW$ be the transition matrix in the new base.  We will have for all $i = 2, \ldots , n$ and $j \in \{0,1\}$:

    \begin{equation}\label{newbaserec1}
    \left[
    \begin{matrix}
    p_j(i)\\
    q_j(i)\\
    r_j(i)
    \end{matrix}
    \right]
    = WL^{i-1}u_j = \lambda_1^{i-1} u_j(1) w_1 + \alpha_{i,j} w_2 + \beta_{i,j} w_3,
    \end{equation}

where $\|(\alpha_{i,j}, \beta_{i,j})\| \leq |\lambda_2|^{i-1} \|u_j(2), u_j(3)\|$.

By a simple calculation the latter implies that

\begin{equation}\label{remest}
\|\alpha_{i,j} w_2 + \beta_{i,j} w_3\| < 0.3409
\end{equation}

 for all $i > 4$ and $j \in \{0, 1\}$.

Let $R_{00}, R_{11}$ be defined as above. Then $|R_{00}| = q_{00}(n)$ and $|R_{11}| = r_{11}(n)$.  In order to calculate $|R_{00} \cup R_{11}|$, we need to find
$|R_{00} \cap R_{11}|$.  The Boolean vectors in $R_{00} \cap R_{11}$ must take the value 1 both on the first and the last coordinates for some inputs $s$
with $s_1 = s_{n+1} = 0$.  This will happen iff $s_2 = s_n = 1$, and we conclude that $|R_{00} \cap R_{11}| = p_{11}(n-2)$.
By the argument preceding Corollary~\ref{cancor} this implies $|R_{00} \cap R_{11}| = {q_00}(n) + p_{11}(n) - p_{11}(n-2)$, and it follows
from~(\ref{newbaserec1}) and~(\ref{remest}) that

\begin{equation}\label{Rest}
|R(n)| = \lambda_1^{n-1}(u_0(1)w_1(2) + u_1w_1(1)) - \lambda_1^{n-3}u_1w_1(1)) + rem(n),
\end{equation}

where $|rem(n) \leq 1.1$ for $n > 6$.

We conclude that

\begin{equation}\label{numest}
\varrho(n) \leq (\lambda_1^{n-3}( \lambda_1^2(u_0(1)w_1(2) + u_1w_1(1)) - u_1w_1(1)) + 1.1)^{1/n},
\end{equation}

for all $n > 6$.

The right hand side of~(\ref{numest}) is less than $10^{1/4}$ for $n > 6$, and Lemma~\ref{rholem} follows by directly calculating
$|R(2)| = 3 = 1.7321^{2}$, $|R(3)| = 5 = 1.7100^{3}$, $|R(4)| = 10 = 1.7783^{4}$, $|R(5)| = 17 = 1.7623^{5}$,
$|R(6)| = 29 = 1.7528^{6}$. $\Box$
\bigskip

\begin{theorem}\label{c22lem}
Suppose there exists a strictly quadratic, bi-quadratic, 0.25-biased $n$-dimensional Boolean system $(\Pi,g)$ with an orbit of size $c^n$.  Then
$c \leq 10^{1/4}$.
\end{theorem}

\noindent
\textbf{Proof:} Let $(\Pi,g)$ be as in the assumption, and let $C$ be an orbit of size $c^n$. Let
$\varrho = \max\{\varrho(k): \ 2 \leq k \leq n\}$.  Call a subset $I \subset [n]$ \emph{closed} if
there exists a subset $J(I) \subset [n]$ with $|J(I)| = |I|$ such that each $g_j$ for $j \in J$ takes inputs from $I$, and call $I$ \emph{minimal closed} if no proper subset of $I$ is closed. Since each node $i$ has both in- and outdegree 2, $[n]$ is the union of pairwise disjoint minimal closed sets $I_k$, for $k \in [K]$, with $2\leq  |I_k| \leq n$ for all $k$ and $J(I_k) \cap J(I_{k'}) = \emptyset$ for $k \neq k'$.
Note that $\sum_{k=1}^K |I_k| = n$.  For each~$k$ the vector of regulatory functions $(g_{1,k}, \ldots , g_{|I_k|, k})$ for elements of $J(I_k)$ forms a minimal Boolean $|I_k|$-block.

If $s \in C$, then $s = g(r)$ for some $r \in C$, and the restriction of $s$ to $J(I_k)$ must be in the range of $(g_{1,k}, \ldots , g_{|I_k|, k})$.  It follows that
$$c^n = |C| \leq \prod_{k=1}^K R(|I_k|) \leq \prod_{k=1}^K \varrho^{|I_k|} = \varrho^n,$$
and the theorem is a consequence of Lemma~\ref{rholem}. $\Box$

\begin{corollary}\label{newupper}
$c(0.25, \alpha, 2, 2) \leq 10^{(2-\alpha)/4}$ for all $\alpha \in [0,1]$.
\end{corollary}

\noindent
\textbf{Proof:}  Suppose $(\Pi, g)$ is a  with at least $\alpha n$
 strictly quadratic
regulatory functions and an orbit of length $c^n$ for some $c > 1$.  By Lemma~\ref{upperlem}, there exists a strictly quadratic, bi-quadratic, 0.25-biased, $(2-\alpha)n$-dimensional Boolean network $(\Sigma, f)$ that also has an orbit of length $c^n$.  By Theorem~\ref{c22lem}, we must have $c^n \leq (10^{1/4})^{(2-\alpha)n}$, and the result follows. $\Box$
\bigskip

The upper bound for $c(0.25, \alpha, 2, 2)$ of Corollary~\ref{newupper}
is less than our previous upper bound $C(0.25, \alpha, 2, 2)$ for $0.7987 < \alpha \leq 1$, but becomes meaningless for $\alpha < 0.7959$, since $(10^{1/4})^{(2-0.7959)} = 2$.  Note also that for $0< \alpha < 1$, both upper bounds for $c(0.25, \alpha, 2, 2)$ exceed the lower bound
 $10^{\alpha/4}2^{(1-\alpha)} \leq c(0.25, \alpha, 2, 2)$ given by Proposition~\ref{1(iii)prop}.

While we believe that all upper bounds $C(\eps, \alpha, b, r)$ substantially overestimate $c(\eps, \alpha, b, r)$ we want to point out that the method of the proof of Corollary~\ref{newupper} does not easily generalize to cases where $b \neq 2$ or $r \neq 2$.

\section{Appendix D: $p$-Fluid and $p$-Unstable Turing Systems}

\begin{definition}\label{fluiddef}
Let $s(0)$ be an initial state of a Boolean system.  We say that the $i$-th variable is \emph{eventually frozen} for the initial state $s(0)$
if $s_i(t)$ takes one of the values $0,1$ only finitely often along the trajectory of $s(0)$.

Let $p \in (0,1]$ and let $k$ be a nonnegative integer.
 A Boolean system is \emph{$p$-fluid} if with probability at least $p$ a randomly chosen initial state has a proportion of at most $p$ eventually frozen variables.

 A Boolean system is \emph{$p$-unstable} if a random single-bit flip in a randomly chosen initial state  moves the trajectory into the basin of attraction of a different attractor with probability at least $p$.
\end{definition}

Here we prove that Theorem~\ref{theorem counterexample}(i) can be strengthened as follows:

\begin{theorem}\label{thm+}  Let $c, p$ be positive constants with $c < 2$, $p < 1$.  Then for all sufficiently large $n$ there exist
$n$-dimensional cooperative Boolean networks that are simultaneously
bi-quadratic, $c$-chaotic, $p$-fluid, and $p$-unstable.
\end{theorem}

\noindent
\textbf{Sketch of the proof:} Let $p, c$ be as in the assumption, and let $(\Pi, g)$
 be an $n$-dimensional Boolean system as constructed in the proof of Theorem~\ref{theorem counterexample}(i).  As in Figure~\ref{figure cooperative}, let $S_1, \ldots , S_{N+1}$ denote the blocks of the
(extended) system, with $S_i(t+1) = S_{i+1}(t)$ for $i \leq N+1$ and
$S_{N+1}(t+1) = S_1(t)$ or $S_{N+1}(t+1) = g(S_1(t))$ depending on the value of the internal variable $mode$.  Then every proper initial state of the system (as defined in the proof of Lemma~\ref{teo cooperative cycle}) is contained in an orbit of length at least $c^n$.

However, most initial states are not proper.
The idea of the proof of Theorem~\ref{thm+} is to modify the system in such a way that in most initial states there will be a sufficiently small $t_0$ such that we will have $S_{N+1}(t+1) = S_1(t)$ for all $t \geq t_0$.  We can accomplish this by modifying the subnetwork $D$ of Figure~\ref{figure cooperative} so that in addition it will have a new pair of output variables $(r_1, r_2)$.  These will be computed as
$r_1 (t+1) = r_1(t) \wedge r_1^*(t)$ and $r_2 (t+1) = r_2(t) \vee r_2^*(t)$, where $r_1^*(\mu-1) = 0$ iff $p(0)$ codes a subset of size less than
$L/2$ and $r_2^*(\mu-1) = 1$ iff $p(0)$ codes a subset of size larger than
$L/2$ (see Figure~\ref{figure cooperative}).  All internal variables of the modified system $D^+$ that are used in the computation of $r_1^*$ and $r_2^*$ will be different from the variables of the original system $D$.  Finally, modify~(\ref{cooperative Turing}) so that $d_1(t) = q_1(t-1)\wedge r_1(t-1)$ and $d_2(t) = q_2(t-1)\vee r_2(t-1)$.

Let us define a \emph{proper initial state} of the modified system as in the proof of Lemma~\ref{teo cooperative cycle}, but requiring in addition that $r_1(0), r_1^*(0)$, and all internal variables of $D^+$ used in the computation of $r^*_1$ are set to 1, and
$r_2(0), r_2^*(0)$, and all internal variables of $D^+$ used in the computation of $r^*_2$ are set to 0.  In this case $r_1(t) = 1$ and $r_2(t) = 0$ throughout the trajectory of any proper initial state, so the modification has no effect on $d(t)$, and the proof of Lemma~\ref{teo cooperative cycle} remains otherwise unchanged.  Thus the modified system will remain $c$-chaotic.

Now suppose the modified system encounters a block $S_{m+\mu + 2}$ with fewer than $L/2$ ones.  Then $r_1$ will take the value 0 at most $\mu$ time steps later, and will stay 0 throughout the trajectory.  Similarly, if the modified system encounters a block $S_{m+\mu + 2}$ with more than $L/2$ ones, then $r_2$ will take the value 1 at most $\mu$ time steps later, and will stay 1 throughout the trajectory.  Thus if the system encounters both a block with fewer than $L/2$ ones and a block with more than $L/2$ ones, then we will have a time $t_0$ with $d(t) = (0,1)$ for all times $t \geq t_0$.  More precisely, let $E_1$ be the event that there are $i, i'$ with $m+\mu+2 \leq i,i' \leq j$ such that $S_i(0)$ has fewer than $L/2$ ones and $S_{i'}(0)$ has more than $L/2$ ones.  If an initial state $S$ belongs to $E_1$, then $S_{N+1}(t+1) = S_1(t)$ for all $t \geq j$.  Let $j$ be large enough so that the probability of $E_1$ is at least
$\sqrt{p}$.

Moreover, let $S_i = (s_i^1, \dots , s_i^L)$ be listed in such an order that $s_i^k(t+1) = s_{i+1}^k(t)$.  Let $E_2$ be the event that for each $k \in [L]$ there exist $j < i, i' \leq N+1$ with $s_i^k(0) \neq s_{i'}^k(0)$.  For sufficiently large $n$, the probability of $E_2$ is at least $\sqrt{p}$.  The events $E_1$ and $E_2$ are independent, thus for sufficiently large $n$, a proportion of at least $p$ of the initial states belong to $E_1 \cap E_2$. It is easy to see that none of the nodes in $S_1 \cup \dots \cup S_{N+1}$ will be eventually frozen for any initial state in
$E_1 \cap E_2$. Since the size of $B$ and $D^*$ depends only on $L$, for sufficiently large $n$ we will have $\frac{|S_1 \cup \dots \cup S_{N+1}|}{n} \geq p$, and $p$-fluidity follows.

In order to prove $p$-instability, we need to consider the probability space of pairs $(S,k)$, where $S$ is a random initial state and $k$ is the position at which the single-bit flip occurs.  Let $j, E_1$ be defined as before, let $E_1^*$ be the event that the first coordinate $S \in E_1$, and let $E_3^*$ be the event that the single-bit flip occurs in some block $S_i$ with $i > j$.    If $(S(0), k) \in E_1^* \cap E_3^*$ and $S^*(0)$ is the initial state obtained from $S(0)$ by the single-bit flip at position $k \in S_i$, then $|S_i((N+1)t) \Delta  S^*_i((N+1)t)| = 1$ for all $t \geq 0$, and  $S_{i'}((N+1)t) = S^*_{i'}((N+1)t)$ for all $t \geq 0$ and $i \neq i'$.  Thus
the single-bit flip moves the system to a different basin of attraction. The events $E_1^*$ and $E_3^*$ are independent.  For sufficiently large $n$ we will have $Pr(E_3^*) > \sqrt{p}$ and thus
$Pr(E_1^* \cap E_3^*) > p$, and $p$-instability follows. $\Box$

\section{Appendix E: $p$-Unstable Strictly Quadratic Systems}

Recall that a Boolean system is \emph{$p$-unstable} if a random single-bit flip in a randomly chosen initial state  moves the trajectory into the basin of attraction of a different attractor with probability at least $p$.  Here we prove two results on such systems.

\begin{proposition}\label{instex}
Let $n$ be a positive integer.  Then there exists a $1$-unstable, strictly bi-quadratic cooperative Boolean system of dimension $2n$.
\end{proposition}

\noindent
\textbf{Proof:}  We construct a $2n$-dimensional Boolean system $(\Pi,g)$ by defining, for $i \in [n]$, the regulatory functions as follows:

\begin{equation}\label{regeqn}
\begin{split}
g_{2i} &= s_{2i - 1} \vee s_{2i},\\
g_{2i-1} &= s_{2i - 1} \wedge s_{2i}.
\end{split}
\end{equation}

Clearly, the resulting system is strictly bi-quadratic and cooperative.  Now consider an initial state $s(0)$ of the system and let $i \in [2n]$.
Note that our choice of the regulatory functions ensures that the number of 1s in the set $\{s_{2i-1}(1), s_{2i}(1)\}$ is the same as the number of 1s in the set $\{s_{2i-1}(0), s_{2i}(0)\}$.  Thus the total number of 1s in $s(0)$ is preserved throughout the trajectory of $s(0)$.  Since the number of 1s changes if we flip a single bit,  each one-bit flip in every initial state moves the system to a different attractor. $\Box$

\begin{theorem}\label{pcthm}
Let $c$ be a constant such that $2 \sqrt{0.75} < c < 2$ and let $p > 0.75 +  \frac{\ln (0.5c)}{2\ln 0.75}$.  Then no strictly quadratic cooperative Boolean system can simultaneously be $c$-chaotic and $p$-unstable.
\end{theorem}

\noindent
\textbf{Proof:} Let $c$ be as in the assumption and assume $(\Pi, g)$ is a $c$-chaotic strictly quadratic cooperative Boolean system of dimension $n$.   A pair $(j, j')$ with $j, j' \in [n]$ will be called \emph{dominating} if there are $i, i' \in [n]$ such that $g_j = s_{i} \vee s_{i'}$ and
$g_{j'} = s_{i} \wedge s_{i'}$.  Let $I$ be the set of all $s_i$ with outdegree 2 that act as input of a dominating pair.

\begin{lemma}\label{dominatinglem}
The set $I$  has cardinality at most $\frac{2n\ln (0.5c)}{\ln 0.75}$.
\end{lemma}

\noindent
\textbf{Proof:} Let $J$ be the union of all dominating pairs for which at least one input variable is in~$I$.  Note that if $(j, j')$, $(k, k')$ are dominating pairs of variables whose input variables contain variables with outdegree 2, then
$\{j, j'\} \cap \{k, k'\} = \emptyset$.  Thus $|J| \geq |I|$.  Moreover, note that if $s$ is a state in an attractor and $(j, j')$ is a dominating pair, then $s_j \geq s_{j'}$; this is our reason for choosing the name `dominating pair.'  It follows that each orbit of the system can have length at most $3^{|J|/2}2^{n-|J|}$.  Since we assumed that there exists an orbit of length at least $c^n$, we must have
$3^{|J|/2}2^{n-|J|} \geq c^n$,
and the lemma follows by taking logarithms. $\Box$
\bigskip

Now let $I_0$ denote the set of $i$ with outdegree zero, $I_1$ the set of $i$ with outdegree 1, and $I_2$ the set of $i$ outside of $I$ with outdegree 2, and let $I_{\geq 3}$ denote the set of odes with outdegree larger than~$2$. Since the sum of all outdegrees must equal the sum of all indegrees and the system was assumed to be strictly quadratic, we must have

\begin{equation*}\label{outdegrees}
|I_1| + 2|I_2| + 2|I| + 3|I_{\geq 3}| \leq 2n = 2(|I_0| + |I_1| + |I_2| + |I| + |I_{\geq 3}|),
\end{equation*}

and hence

\begin{equation*}
|I_{\geq 3}| \leq 2|I_0| +  |I_1|, 
\end{equation*}

which gives us 

\begin{equation}\label{degreetotals}
3|I_0| + 2|I_1| + |I_2| \geq n - |I| \geq n - \frac{2n\ln (0.5c)}{\ln 0.75}.
\end{equation}

Now consider a random initial state $s(0)$ and the state $s^*(0)$ obtained by flipping the value of the variable $s_i(0)$.  Clearly, if
$i \in I_0$, then $s(1) = s^*(1)$, since the variable $s_i$ is not used at all to calculate the next state.  In particular, a single-bit flip of a single variable in $I_0$ will leave the system on the same trajectory with probability~1.  If $i \in I_1$, then there is exactly one $s_j$ for which $s_i$ acts as input.  Assume wlog that $g_j = s_i \vee s_{i'}$; the case of the conjunction is analogous.  Note that $i' \neq i$ since the system was assumed strictly quadratic.   With probability 0.5, we will have $s_{i'}(0) = s^*_{i'}(0) = 1$.   In this case the bit flip has no effect on the value of $s_j(1)$ and again we get $s(1) = s^*(1)$.  In particular, a single-bit flip at $s_i$ for $i \in I_1$ will leave the system on the same trajectory with probability at least 0.5.

Now consider the case when $i \in I_2$.  Then there are $j \neq j'$ such that $s_i$ acts as input to both $g_j$ and $g_{j'}$.
Then $g_j = s_i \, L \, s_{i'}$ and $g_{j'} = s_i \, K \, s_{i''}$, where $L, K$ stand for the possible logical operators $\vee, \wedge$.
First assume $i' = i''$.  Then we must have $L = K$, otherwise the pair $(j, j')$ or the pair $(j',j)$ would be dominating, which possibility we have excluded by making $I_2$ disjoint from $I$.  But if $L = K$, then the exact same argument as for $i \in I_1$ shows that with probability 0.5 the bit flip at $s_i$ has no effect on the successor states $s_j(1), s_{j'}(1)$ and hence on the trajectory of $s(0)$.

Finally, assume $i' \neq i''$ and wlog that $g_j = s_i  \vee  s_{i'}$ and $g_{j'} = s_i \wedge s_{i''}$.  Then with probability 0.25, we will have $s_{i'}(0) = 1$ and $s_{i''}(0) = 0$, in which case the single-bit flip at $s_i$ has no effect on the successor state and
$s(1) = s^*(1)$.

From the above and~(\ref{degreetotals}) we conclude that

\begin{equation*}\label{flipprob1}
Pr(s(1) = s^*(1)) \geq \frac{|I_0| + 0.5|I_1| + 0.25|I_2|}{n} \geq 0.25 - \frac{\ln (0.5c)}{2\ln 0.75}.
\end{equation*}

Since the single-bit flip cannot move the system to a different basin of attraction unless $s(1) \neq s^*(1)$, and the theorem follows.
$\Box$

\section{Appendix F: Connection with Boolean Delay Systems}

One can interpret Theorem~\ref{GenTthm} in a different way.
Variables with monic regulatory functions in cooperative Boolean systems just record the values of other variables at some time in the past (less than $n$ steps earlier).  Thus if we allow time delays in the definitions of regulatory functions, we can remove all but the
first variable on each `tape' and define a \emph{Boolean delay system} on the remaining variables that will have equivalent dynamics, in particular, that will have orbits of the same length as the original system.

Continuous-time Boolean delay systems were studied in \cite{GhilI, GhilII}; see \cite{GhilIII} for a comprehensive survey and additional references.  In this framework, time $t$ takes positive real numbers as values, and an $m$-dimensional Boolean System is defined by regulatory
functions $f_i: \{0,1\}^m \rightarrow \{0, 1\}$ for $i \in [m]$ such that

\begin{equation}\label{delayeqn}
\begin{split}
s_1(t) &= f_1(s_1(t-t_{11}), s_2(t-t_{12}), \ldots , s_1(t-t_{1m})),\\
s_2(t) &= f_2(s_1(t-t_{21}), s_2(t-t_{22}), \ldots , s_1(t-t_{2m})),\\
&.\\
&.\\
&.\\
s_m(t) &= f_n(s_1(t-t_{m1}), s_2(t-t_{m2}), \ldots , s_1(t-t_{mm})),
\end{split}
\end{equation}
where the $t_{ij}$'s are positive time delays.

Now suppose $(\Pi, g)$ is an $n$-dimensional discrete Boolean system, and $s_k$ is a variable with a monic regulatory function $g_k$.
Then either their exists a sequence of variables $k_1, \ldots , k_\ell = k$ such that $g_{k_{i+1}} = s_{k_i}$ or $g_{k_{i+1}} = \neg s_{k_i}$ for all
$i \in [\ell -1]$ and either $g_{k_1}$ is non-monic or $k_1 = k_\ell$.  The sequence $k_1, \ldots , k_\ell$ is uniquely determined by $s_k$, we will
call it the \emph{tape of~$k$.}  If $k_1 = k_\ell$, then we will say that \emph{$s_k$ has a read-only tape.}

Now assume for simplicity of notation that the set of variables numbered $s_1 , \ldots , s_m$ comprises all variables with non-monic regulatory
functions, together with exactly one variable from each read-only tape. Then for each variable $s_k$ of $(\Pi, g)$ with  $k > m$ there exists
$i(k) \in [m]$ such that for all states in any attractor of $(\Pi, k)$ we either have  $s_k (t) = s_{i(k)}(t-\ell(k))$ for all $t$  or
$s_k (t) = \neg s_{i(k)}(t-\ell(k))$.  In the former case, we call $s_k$ is a \emph{direct memory variable;} in the latter case we call $s_k$ an
\emph{inverted memory variable.} The number $\ell(k)$ represents the length of the tape for $s_k$ if the tape is not read-only, and has a similar interpretation for read-only tapes.  We can also interpret $\ell(k)$ as a time delay, and define an $m$-dimensional Boolean delay system $(\Sigma, f)$ as follows: For each $i \in [m]$, let

\begin{equation}\label{fidefn}
f_i(t) = g_i(s_1(t-1), \ldots , s_m(t-1), \hat{s}^{m+1}_{i(m+1)}(t - \ell(m+1) - 1), \ldots , \hat{s}^n_{i(n)}(t - \ell(n) - 1)),
\end{equation}
where $\hat{s}^k_{i(k)} = s_{(t - \ell(k) - 1)}$ if  $s_k$ is a direct memory variable and
and $\hat{s}^k_{i(k)} = \neg s_{(t - \ell(k) - 1)}$ if  $s_k$ is an inverted memory variable.

Let $L$ be the maximum delay in $(\Sigma , f)$.  An initial state in this system is given by specifying the values of
$s_i(t)$ for all $i \in [m]$ and $t \in [0,L)$.  If we choose the initial state in such a way that all $s_i(t)$'s are constant on every interval
$[\ell-1, \ell)$
for $\ell \in [L]$, then all $s_i(t)$'s will remain constant on intervals $[\ell-1, \ell)$ for all positive integers $\ell$ throughout the trajectory of this initial state.

Now consider an initial state $(r_1(0), \ldots , r_n(0))$ of $(\Pi, g)$, and define an initial state $(s_1, \ldots , s_m)\upharpoonright[0,L)$ of
$(\Sigma , f)$ so that $s_i(t) = r_i(\ell-1)$ whenever $\ell \in [L]$ and $\ell-1 \leq t < \ell$.  It is straightforward to verify that
for all integers $\ell \geq L$ and $\ell \leq t < \ell + 1$ and $i \in [m]$ we will have $s_i(t) = r_i(\ell)$.  Thus the dynamics of the systems
$(\Pi, g)$ and $(\Sigma, f)$ will be equivalent along all orbits in an obvious sense, and we will simply write that the systems
$(\Pi, g)$ and $(\Sigma, f)$ are \emph{equivalent.}

We get the following Corollary of Theorem~\ref{GenTthm}.

\begin{corollary}\label{delaycorol}
For any given
 $\alpha, \eps > 0$ and positive integers $b,r$ there exists a positive constant  $c < 2$ such that for sufficiently large $n$, every $c$-chaotic $n$-dimensional $\eps$-biased $(b,r)$-Boolean system is equivalent to a Boolean delay system with integer delays and at most $\alpha n$ Boolean variables.
\end{corollary}

\noindent
\textbf{Proof:} Let $\alpha, \eps$ be as in the assumptions and let $(\Pi, g)$ be an $n$-dimensional $\eps$-biased $(b,r)$-Boolean system with an orbit of length at least $c^n$, where $c > c(\eps, \alpha/2, b, r)$.  By Theorem~\ref{GenTthm}, $(\Pi, g)$ is an $(\frac{\alpha}{2} n, n)$-Turing system.  By the argument above, there exists an equivalent Boolean delay system $(\Sigma, f)$ with integer delays whose dimension $m$ is equal to the sum of the number of non-monic regulatory functions in $(\Pi, g)$ and the number of read-only tapes in $(\Pi, g)$.  It remains to show that
the latter number cannot exceed $\frac{\alpha}{2} n$ if $c$ is sufficiently close to~2.

Let us define the \emph{read-only part} or \emph{strictly monic part} $R(g)$ of a
Boolean system $(\Pi, g)$ as the union of all its  read-only tapes.  Now Corollary~\ref{delaycorol} is a consequence of the following observation if we let $\delta = \frac{\alpha}{2}$.

\begin{lemma}\label{readonlythm}
Let $\delta > 0$ and let $c > 2^{1-\delta}$.  Then for sufficiently large $n$, no $n$-dimensional Boolean system $(\Pi, g)$ with $|R(g)| \geq \delta n$ can be $c$-chaotic.
\end{lemma}

\noindent
\textbf{Proof:} First note that if $T = \{k_1, \ldots , k_m\}$ is a read-only tape of length $m$, then the dynamics of the system on $T$ is can be described by cyclical shifts, possibly with negations in some positions.  Thus given any initial state $s(0)$ of the system, the vector
$(s_{k_1}(t), \ldots , s_{k_m}(t))$ can take  at most $2m$ distinct values throughout the trajectory of $s(0)$.
Since $R(g) = \{\ell_1, \ldots , \ell_{R}\}$ is the union of pairwise disjoint read-only tapes $T_1, \ldots , T_v$ with
$|T_1| + \dots +|T_v| = |R(g)|$, it also follows from the same observation that
 the vector $(s_{\ell_1}(t), \ldots , s_{\ell_{R}}(t))$ can take  at most $2lcm \left(\{|T_1|, \ldots , |T_v|\}\right)$ distinct values throughout the trajectory of $s(0)$, where $lcm$ stand for the least common multiple.

Let $P(N)$ denote the maximum value of $lcm\left( \{m_1, \ldots , m_r\}\right)$ with $\sum_{i=1}^r m_i = N$.
Then
$P(N)= e^{ \sqrt{N \ln N} (1+o(1)) }$  as $N\to \infty$ (see Chapter 13 of \cite{Landau}). It follows that for any given initial state $s(0)$, the vector of values of the variables in $R(g)$ can take at most $2e^{ \sqrt{|R(g)| \ln |R(g)|} (1+o(1)) }$ different values in any orbit.
 Thus under the assumptions of the lemma, the size of any orbit is bounded from above by $2^{(1 - \delta)n}2e^{\sqrt{n \ln n}(1 + o(1)}$, which is less than $c^n$ for $c$ as in the assumption and sufficiently large~$n$. $\Box$ $\Box$

\end{document}